\documentclass[aip,preprint,preprintnumbers,amsmath,amssymb]{revtex4}

\usepackage{graphicx}
\usepackage{dcolumn}
\usepackage{bm}

\usepackage[utf8]{inputenc}
\usepackage[T1]{fontenc}
\usepackage{mathptmx}
\usepackage{etoolbox}
\usepackage{epsfig}
\makeatletter
\def\@email#1#2{%
 \endgroup
 \patchcmd{\titleblock@produce}
  {\frontmatter@RRAPformat}
  {\frontmatter@RRAPformat{\produce@RRAP{*#1\href{mailto:#2}{#2}}}\frontmatter@RRAPformat}
  {}{}
}%
\makeatother

\newcommand{\mE}{\mathcal{E}}
\newcommand{\mA}{\mathcal{A}}
\newcommand{\mH}{\mathcal{H}}

\newcommand{\rf}{Ref.~\onlinecite}
\newcommand{\cp}{Ref.~\onlinecite{benisti20I}}
\newcommand{\kld}{k\lambda_D}
\newcommand{\mL}{\mathcal{L}}
\newcommand{\mF}{\mathcal{F}}
\newcommand{\mk}{\bm{k}}

\newcommand{\wl}{\omega_{\rm{las}}}

\usepackage{color} 

\usepackage{hyperref}
\hypersetup{
    colorlinks=true,
    citecolor=blue,
    linkcolor=blue,
    filecolor=blue,
    urlcolor=blue,
}

\usepackage{textcomp}

\begin {document}
\title{Nonlinear adiabatic electron plasma waves. II. Applications.}
\author{D. B\'enisti}
\email{didier.benisti@cea.fr}
\author{D.F.G. Minenna}
\author{M. Tacu}
\author{A. Debayle}
\author{L. Gremillet}
\affiliation{ CEA, DAM, DIF F-91297 Arpajon, France and Universit\'e Paris-Saclay, CEA, LMCE, 91680 Bruy\`eres-le-Ch\^atel, France.}
\date{\today}
\begin{abstract}
In this article, we use the general theory derived in the companion paper [M. Tacu and D. B\'enisti, Phys. Plasmas (2021)] in order to address several long-standing issues regarding nonlinear electron plasma waves (EPW's). First, we discuss the relevance, and practical usefulness, of stationary solutions to the Vlasov-Poisson system, the so-called Bernstein-Greene-Kruskal modes, to model slowly varying waves. Second, we derive an upper bound for the wave breaking limit of an EPW growing in an initially Maxwellian plasma. Moreover, we show a simple dependence of this limit as a function of $k\lambda_D$, $k$ being the wavenumber and $\lambda_D$ the Debye length. Third, we explicitly derive the envelope equation ruling the evolution of a slowly growing plasma wave, up to an amplitude close to the wave breaking limit. Fourth, we estimate the growth of the transverse wavenumbers resulting from wavefront bowing by solving the nonlinear, nonstationary, ray tracing equations for the EPW, together with a simple model for stimulated Raman scattering. 
\end{abstract}
\maketitle
\section{Introduction}
\label{I}
Although electron plasma waves (EPW's) have been extensively studied since the seminal work by Tonks and Langmuir~\cite{langmuir}, a complete nonlinear theory for these waves is still to be derived. Actually, this  remains a formidable task even when one restricts to a kinetic description in the classical regime. Indeed, this would require a theoretical resolution of the Vlasov-Maxwell equations, valid whatever the space and time variations of the wave and of the plasma. In this article, we do not aim at such a universality. Instead, we focus on a particularly important class of nonlinear EPW's, the so-called adiabatic ones. These mainly result from the electron motion, provided that this motion may be accurately described by making use of the adiabatic approximation, i.e., by assuming that the dynamical action remains essentially constant (up to some geometrical changes entailed by separatrix crossing). 
As discussed in~\cp, this lets us restrict to waves such that $\gamma/kv_{th}\alt0.1$, where $\gamma$ is the typical wave growth rate, $k$ is the wavenumber and $v_{th}$ is the electron thermal velocity. Moreover, we also restrict to propagating waves, so that physics situations which could lead to Anderson-like localization~\cite{doveil} are excluded. Under these conditions, we address in this article several long-standing issues regarding nonlinear EPW's. 

Fist of all, there has been a considerable effort to derive stationary solutions to the Vlasov-Poisson system, which are the so-called Bernstein-Greene-Kruskal (BGK) modes~\cite{bgk}. However, since a wave is never exactly stationary and an EPW is never exactly electrostatic~\cite{benisti20I}, the relevance of BGK modes to model actual physics problems is not always clear. In particular, one may wonder whether these modes may correctly approximate slowly growing waves, resulting from an instability, and described in the companion paper~\cite{benisti20I}. We address this issue in Section~\ref{II}, where we compare the electrostatic field of previously proposed BGK modes with that derived in~\cp. This lets us discuss when an accurate description of the electrostatic field may be obtained much more rapidly and more simply than by going through the whole derivation of Ref.~\onlinecite{benisti20I}. In this respect, special attention is paid to the well-known solution provided by Dawson in Ref.~\onlinecite{dawson}. Moreover, in Section~\ref{II}, we clearly explain the analogies and differences between our theory and the derivation of BGK modes. 

Second, a nearly monochromatic wave cannot grow beyond a maximum amplitude known as the wave breaking limit. Deriving this limit is a long-standing and important issue. Indeed, this would allow to conclude about the saturation level of an instability, or about the effectiveness of stimulated Raman scattering (SRS) as a means for laser pulse amplification~\cite{malkin}. One way to obtain an upper bound for the wave breaking limit is to find the maximum amplitude allowing a solution to the nonlinear dispersion relation. This is what we do in Section~\ref{III} using the dispersion relation derived in the companion paper~\cite{benisti20I}. Moreover, we compare our results with those obtained by Coffey in Ref.~\onlinecite{coffey} for a stationary wave in an initially waterbag distribution function.  Furthermore, we discuss the relevance of the upper bound thus derived. 

Third, in order to fully describe a nonlinear EPW, one must be able to predict the space and time evolution of its amplitude. Resorting to envelope equations has proven to be a  very effective and accurate way to do so for slowly varying waves \cite{brama,benisti10,benisti12,benisti18}. Such equations have been derived in Refs.~\onlinecite{dodin1} and \onlinecite{benisti16} within the geometrical optics limit and by assuming a near adiabatic electron motion. They are valid whatever the harmonic content of the wave which is, however, not specified. Consequently, no explicit analytical formula is provided, except in Ref.~\onlinecite{benisti16} when the electrostatic field is assumed to be sinusoidal (but without discussing the range of validity of the sinusoidal approximation). Using the discussion of Section~\ref{II} regarding the relevance of BGK modes, we provide in Section~\ref{IV} explicit expressions for the nonlinear envelope equation of growing electron plasma waves, which are accurate whatever $\kld$ ($k$ being the wavenumber and $\lambda_D$ the Debye length), and up to amplitudes close to the wave breaking limit. 

Fourth, an EPW, strongly driven into the nonlinear regime by SRS from a laser hot spot, exhibits large transverse wavenumbers. These have been evidenced experimentally in \rf{rousseaux} using Thomson scattering, and shown to be much larger than expected from the opening angle of the focal spot. Now, there may be two different reasons for the growth of these transverse modes. They may result from an instability due to trapped particles, as shown numerically in Refs.~\onlinecite{rousseaux,masson,berger,silva}. They may also be due to wavefront bowing, observed numerically in Refs.~\onlinecite{rousseaux,masson,berger,silva,yin07,yin08,yin,ban11,yin12,yin13}. Indeed, an SRS-driven EPW grows faster where the laser intensity is larger, near the center of the focal spot. Consequently, the wave amplitude is inhomogeneous in the direction transverse to the laser propagation. Then, so are the wave frequency and wave phase velocity, since these are nonlinear functions of the amplitude~\cite{benisti20I,benisti08}.  As a result, the wavefront bends, usually so as to induce self-focussing~\cite{rousseaux,masson,berger,silva,yin07,yin08,yin,ban11,yin12,yin13}. This, in turn, entails the growth of transverse modes, since the local wavenumber is perpendicular to the wavefront. Kinetic simulations, either using  a particle-in-cell (PIC) or a Vlasov code, always show the wavefront bowing and the unstable growth of secondary modes. Consequently, one cannot tell which is the dominant effect, as discussed in detail in Ref.~\onlinecite{rousseaux}. This issue is addressed in Section~\ref{V}, where we calculate the transverse wavenumbers which only result from wavefront bowing. To do so, we clearly need to go beyond the paraxial~\cite{lax,riazuelo}~or quasioptical~\cite{permitin,dodin_ray}~approximations. 
Indeed, we have to solve, very finely, for the time variations of the EPW wavenumber, which depend on the local wave amplitude. In other words, we have to solve the nonlinear, nonstationary, ray-tracing equations for the EPW, together with its envelope equation. Our numerical resolution follows from that introduced in \rf{deb19}, where the physical space is subdivided into regular cells. In order to derive the nonlinear ray dynamics, we need the local value of the EPW amplitude. This is estimated as an average over the rays located within the same cells.  More precisely, using the same technique as that introduced in particle-in-cell (PIC) codes, the EPW amplitude is first estimated on the cell nodes by making use of a shape factor. Then, it is projected back onto the rays by resorting to the same shape factor. For this reason, we dubbed our numerical scheme ``ray-in-cell'' (RIC).  
By comparing the results of our model with those from two-dimensional (2-D) PIC simulations of SRS, we can conclude on the ability to derive the EPW transverse spectrum by relying, only, on wavefront bowing. This is an important issue because the opening angle of the backscattered light directly follows from that of the EPW. Then, a simple model that quantifies the transverse modes of the EPW is needed at least for two reasons: (i) to correctly predict the impact of SRS on the plasma hydrodynamics; (ii) to properly interpret experiments of laser-plasma interaction as regards the direction of the backscattered light.

This paper is organized as follows. In Section~\ref{II}, we compare the electrostatic field derived from the adiabatic theory of the companion paper, \rf{benisti20I}, with those of previously proposed BGK modes. Section~\ref{III} addresses the wave breaking limit for adiabatic EPW's. In Section~\ref{IV}, we provide an explicit expression for the nonlinear envelope equation of a growing electron plasma wave, which is valid whatever $\kld$ and up to amplitudes close to the wave breaking limit. Section~\ref{V} introduces a simple model to quantify the transverse modes resulting from wavefront bowing, and compares the predictions of the model with those of 2-D PIC simulations. Section~\ref{VI} summarizes and concludes our work. 

\section{Comparisons between nonlinear adiabatic plasma waves and BGK modes}
\label{II}
\subsection{Analogies and differences between adiabatic waves and BGK modes}
\label{II.A}
There are clear differences between the adiabatic waves considered in this paper and BGK modes. Indeed, the latter modes are stationary solutions to the Vlasov-Poisson system and, most often, they are space-periodic, so that the mode amplitude is time and space independent. By contrast, although its variations must be slow, the amplitude of an adiabatic wave may vary in space and time. Moreover, BGK modes are purely electrostatic while we showed in~\cp~that nonlinear adiabatic waves had a nonzero vector potential. However, when the vector potential is negligible, for a uniform wave, and for each fixed value of the amplitude, a nonlinear adiabatic EPW, as derived in~\rf{benisti20I}, {\em{is}} a BGK mode.  

Nevertheless, in spite of the previous strong analogy,  our theory is developed in a spirit totally different from that leading to BGK modes. Indeed, usually, nothing is said about the way a BGK mode has, or could have been, generated. Usually, such a mode has no history. By contrast, in~\rf{benisti20I}, we build the self-consistent wave potential and electron distribution function by accounting for the full wave history. In particular, for a given wave amplitude, our result will be different depending on whether the wave has kept on growing or if its amplitude has not been a monotonous function of time. Actually, our theory is designed to predict the space and time evolution of the wave, by using envelope equations like those derived in Section~\ref{IV}. However, the general derivation of~\cp~is quite tedious, while BGK modes are explicit solutions to the Vlasov-Poisson system, which usually depend on several free parameters. Then, one may wonder whether the theory could be simplified by choosing those parameters  so as to get  a relevant description of nonlinear adiabatic  waves. In particular, we discuss in Paragraph~\ref{II.B}~the relevance of the very simple solution introduced by Dawson in~\rf{dawson}, using previous results by Akhiezer and Lyubarskizs~\cite{akhiezer}. Dawson's solution is for nonlinear plane waves in a cold plasma, and it depends on a single parameter, the wave amplitude. The corresponding electric field reads
\begin{equation}
\label{E}
E(x)=\frac{n_ee}{\varepsilon_0}X_0\sin[kx_0(x)],
\end{equation}
with
\begin{equation}
\label{de}
x=x_0+X_0\sin(kx_0),
\end{equation}
where $n_e$ and $-e$ are the electron density and charge. 

\subsection{Detailed comparisons between uniformly growing adiabatic waves and Dawson's solution for nonlinear plane waves in a cold plasma}
\label{II.B}
\subsubsection{Field profile}
\label{II.B.1}
Let us introduce the dimensionless electric field,
\begin{equation}
\label{me}
\mE\equiv (ek/m\omega_{pe}^2)(E-E_0),
\end{equation}
where $m$ is the electron mass, $\omega_{pe}=\sqrt{n_ee^2/\varepsilon_0m}$ is the plasma frequency, and $E_0$ is the space-averaged value of $E(x)$ over one wavelength.

For the electric field, Eq.~(\ref{E}), proposed by Dawson, $E_0=0$ and
\begin{equation}
\label{de}
\mE(x)=\mathfrak{E}\sin[kx_0(x)],
\end{equation}
where $\mathfrak{E}=kX_0$. 

For the adiabatic waves of~\cp, $(E-E_0)$ would just be the electrostatic field. Moreover, $A_0=-\int_0^tE_0(u)du$, as derived in \cp, remains constant when the wave amplitude does not change. This means that, if the adiabatic EPW reaches a given amplitude at $t=t_0$, that does not change whenever $t>t_0$, $E_0=0$ for times larger than $t_0$. Hence, for an adiabatic wave with constant fixed amplitude, $E_0=0$. 
\begin{figure}[!h]
\centerline{\includegraphics[width=12cm]{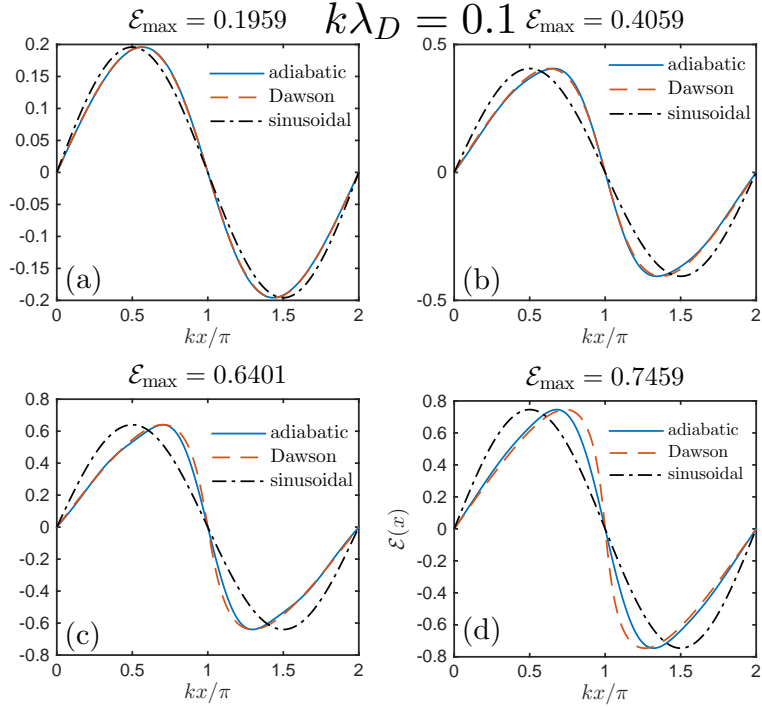}}
\caption{\label{f1} (Color online) Comparisons of the profile of $\mE(x)$ for adiabatic waves (blue solid line) with that proposed by Dawson (red dashed line) and with that of a purely sinusoidal wave (black dashed-dotted line), when $k\lambda_D=0.1$ and ; panel (a), when $\mathcal{E}_{\max}\approx0.1959$ ; panel (b),  when $\mathcal{E}_{\max}\approx0.4059$ ; panel (c),  when $\mathcal{E}_{\max}\approx0.6401$ ; panel (d),  when $\mathcal{E}_{\max}\approx0.7459$.}
\end{figure}
\begin{figure}[!h]
\centerline{\includegraphics[width=12cm]{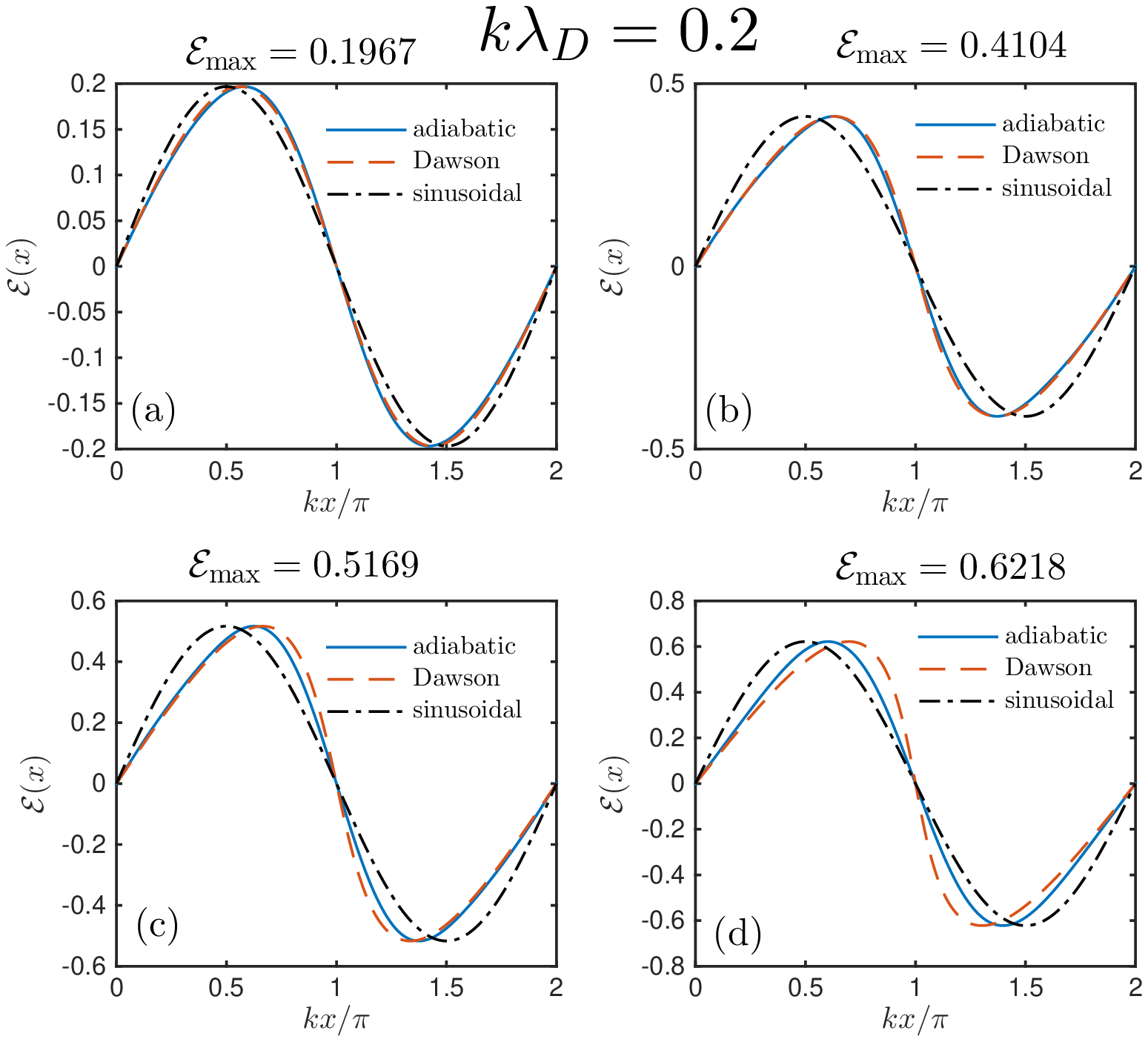}}
\caption{\label{f2} (Color online) Comparisons of the profile of $\mE(x)$ for adiabatic waves (blue solid line) with that proposed by Dawson (red dashed line) and with that of a purely sinusoidal wave (black dashed-dotted line), when $k\lambda_D=0.2$ and ; panel (a), when $\mathcal{E}_{\max}\approx0.1967$ ; panel (b),  when $\mathcal{E}_{\max}\approx0.4104$ ; panel (c),  when $\mathcal{E}_{\max}\approx0.5169$ ; panel (d),  when $\mathcal{E}_{\max}\approx0.6218$.}
\end{figure}

Figs.~\ref{f1}~and~{\ref{f2} compare the profiles of the electric field for adiabatic waves (derived by accounting for harmonics 1 to 3 in the scalar potential) with those of Dawson's solution,~Eq.~(\ref{de}), and with those of a purely sinusoidal wave, for given maximum values, $\mathcal{E}_{\max}$, of the dimensionless field. Fig.~\ref{f1} is for $\kld=0.1$ while Fig.~\ref{f2} is for $\kld=0.2$.

 When $\kld=0.1$, the field profile proposed by Dawson agrees very well with that of nonlinear adiabatic waves whenever $\mathcal{E}_{\max}\alt0.64$. Indeed, if we denote by $\delta \mE_D$ the difference between the electric field derived from Dawson's formula Eq.~(\ref{de}) and that derived from the adiabatic theory, $\sqrt{\langle\delta\mE_D^2\rangle}/\sqrt{\langle\mE^2\rangle}$ is less than 10\% whenever $\mathcal{E}_{\max}\alt0.64$ (it is close to 9\% when $\mathcal{E}_{\max}=0.6401$ and close to 0.5\% when $\mathcal{E}_{\max}=0.1959$). When $\mathcal{E}_{\max}\approx 0.7459$, which is close to the largest amplitude allowing solutions to the adiabatic nonlinear dispersion relation, the field profile proposed by Dawson is slightly steeper than that of adiabatic waves. Indeed, if we denote by $\delta x_D$ (respectively by $\delta x_a)$ the difference between the $x$-position of the minimum and maximum values of Dawson's electric field (respectively of the adiabatic electric field), $k\delta x_D/\pi\approx0.53$ while  $k\delta x_a/\pi \approx0.64$ when $\mathcal{E}_{\max}\approx0.7549$. Nevertheless, whatever the amplitude, the electrostatic field for nonlinear adiabatic waves is better approximated by Dawson's solution than by a sine function.

When $\kld=0.2$ and $\mathcal{E}_{\max}\alt 0.5169$, Dawson's profile for the electrostatic field is very close to that of nonlinear adiabatic waves. Indeed, $\sqrt{\langle\delta\mE_D^2\rangle}/\sqrt{\langle\mE^2\rangle}$ is less than 10\% whenever  $\mathcal{E}_{\max}\alt 0.5169$ (it is close to 10\% when  $\mathcal{E}_{\max}\approx 0.5169$ and close to 2\% when $\mathcal{E}_{\max}\approx 0.1967$). However, when $\mathcal{E}_{\max}\approx 0.6218$, which is close to the maximum amplitude allowing a solution to the nonlinear adiabatic dispersion relation, Dawson's profile is slightly steeper than the adiabatic one, $k\delta x_D/\pi\approx 0.6$ while $k\delta x_a/\pi\approx 0.8$. The sinusoidal profile also provides quite a good approximation of the adiabatic one. Indeed, whenever $\mathcal{E}_{\max}\alt0.6218$,  $\sqrt{\langle\delta\mE_s^2\rangle}/\sqrt{\langle\mE^2\rangle}<20\%$, where $\delta \mathcal{E}_s$ is the difference between the adiabatic and sinusoidal electric fields. Actually, the sinusoidal profile is slightly more accurate than Dawson's one for the largest values of $\mathcal{E}_{\max}$. In particular, when  $\mathcal{E}_{\max}\approx 0.6218$, $\sqrt{\langle\delta\mE_D^2\rangle}/\sqrt{\langle\mE^2\rangle}\approx21\%$ while $\sqrt{\langle\delta\mE_s^2\rangle}/\sqrt{\langle\mE^2\rangle}\approx17\%$. Hence, when $\kld=0.2$, the advantage of resorting to Dawson's profile in order to approximate adiabatic waves, instead of simply using a sine function, is less obvious than when $\kld=0.1$, although Dawson's profile is more accurate whenever $\mathcal{E}_{\max}\alt 0.6$. 

Increasing $\kld$ beyond 0.2 lets nonlinear adiabatic waves get closer and closer to sinusoids. Actually, whenever $\kld>0.3$, they are better approximated by a sine function than by Dawson's profile (not shown here).

In conclusion,  we find that the electrostatic field of uniformly growing adiabatic waves is well approximated by the solution proposed by Dawson whenever $\kld\alt0.2$, although the accuracy decreases close to the wave breaking limit. This result is expected, since Dawson only investigated waves in a cold plasma, i.e., in the limit when $\omega/kv_{th}\rightarrow\infty$, where $\omega$ is the wave frequency and $v_{th}$ the electron thermal speed. Now, in a plasma with finite temperature, and in the linear limit, $\omega/kv_{th}\approx 1/\kld$, so that the cold plasma limit is more relevant for smaller values of $\kld$. However, as the wave amplitude increases, $\omega$ decreases, so that the cold plasma limit becomes less accurate. 

If the wave amplitude does not keep on increasing, some electrons will be detrapped, which would change the distribution function. How this would impact the previous conclusions regarding the relevance of Dawson's solution depends on the variations of $V_\phi=\omega/k-eA_0/m$, $A_0$ being the wave vector potential. If $V_\phi$ changes more slowly than the separatrix width (in velocity), electrons are detrapped symmetrically with respect to $V_\phi$. Then, detrapping would not significantly change the values of $\langle\cos(j\varphi)\rangle$ and, therefore, the harmonics content of the field. In this case, Dawson's solution accurately models the electrostatic field of nonlinear adiabatic waves even when they are not uniformly growing. However, only the theory of~\cp~can address the most general situation, and remains valid whatever variations of $V_\phi$ compared to those of the separatrix width.

\subsubsection{Nonlinear dispersion relation}
\label{II.B.2}
In this Paragraph, we derive an approximate nonlinear adiabatic dispersion relation using Dawson's solution for the electrostatic field, and compare it against the results found from~\cp. More precisely, we still derive $V_\phi=\omega/k-eA_0/m$ by solving~\cite{noteD}
\begin{equation}
\label{dr}
-2\langle\cos(\varphi)\rangle=\Phi_1,
\end{equation}
where $\Phi_1$ is the first harmonic of the dimensionless potential, $\Phi$, such that $\partial_\varphi\Phi=-\mE(\varphi)$. Here, $\mE(\varphi)$ is a plain generalization of Eq.~(\ref{me}), namely,
\begin{equation}
\mE(\varphi)=(ek/m\omega_{pe}^2)[E(\varphi)-E_0(t)],
\end{equation}
where $\varphi$ now depends on space and time, $\partial_x\varphi=k$ and $\partial_t\varphi=-\omega$. As for $\langle\cos(\varphi)\rangle$ in Eq.~(\ref{dr}), it is still given by Eq.~(31)~of~\cp~except that, instead of using the self-consistent potential, we use that derived from Dawson's solution, namely, 
\begin{equation}
\partial_\varphi\Phi=-\mathfrak{E}\sin(\varphi_0),
\end{equation}
where $\varphi_0$ is related to $\varphi$ through
\begin{equation}
\label{phi}
\varphi=\varphi_0+\mathfrak{E}\sin(\varphi_0).
\end{equation}
Then, 
\begin{eqnarray}
\label{9}
\Phi&=&\mathfrak{E}\cos(\varphi_0)-\frac{\mathfrak{E}^2}{2}\sin^2(\varphi_0),\\
\label{phi1d}
\Phi_1&=&\mathfrak{E}[J_0(\mathfrak{E})-J_2(\mathfrak{E})]+\frac{\mathfrak{E}^2}{2}[J_1(\mathfrak{E})+J_3(\mathfrak{E})],
\end{eqnarray}
where $J_n(\mathfrak{E})$ is the Bessel function of order~$n$~\cite{abramowitz}. Hence, Eq.~(\ref{dr}) can be sloved without having to self-consistently calculate the wave potential, which considerably simplifies the derivation of $V_\phi$. Moreover, like in~\cp,  we impose the conservation of the total electron momentum, and derive the nonlinear wave frequency from Eq.~(32) of~\cp. 

Now, it is clear that Eq.~(\ref{9}) for $\Phi$ is not the exact potential of nonlinear adiabatic waves. Consequently, using Dawson's potential would only yield approximate values for $\omega$, which we henceforth denote by $\omega_D$. However, from the discussion of Paragraph~\ref{II.B.1}, these are expected to be accurate. We check the accuracy of $\omega_D$ by comparing their values against those of $\omega_3$, derived from the adiabatic theory of~\cp~by accounting for three harmonics in the potential, and with those of $\omega_1$, derived by assuming a sinusoidal potential. More precisely, we compare the values of $\vert\omega_3-\omega_D\vert$ with those of $\vert\omega_3-\omega_1\vert$ and those of $\vert\delta\omega\vert$, where $\delta\omega$ is the nonlinear frequency shift calculated as $\delta\omega=\omega_3(\Phi_1)-\omega_3(\Phi_1=0)$. Fig.~\ref{f3} shows such comparisons as a function of $\kld$ when $\Phi_1=0.2$, $\Phi_1=0.3$, $\Phi_1=0.4$ and $\Phi_1=0.5$, while Fig.~\ref{f3b} plots such comparisons as a function of $\Phi_1$ when $\kld=0.1$, $\kld=0.15$, $\kld=0.2$ and $\kld=0.3$. From these figures, we can now discuss when approximate values of the nonlinear frequency may be considered as accurate, i.e., when they differ from $\omega_3$, which is our reference, by much less than $\vert\delta\omega\vert$. Only when this condition is fulfilled may the approximate values for the nonlinear frequency be used to derive an accurate nonlinear ray tracing, as that described in Section~\ref{V}.

\begin{figure}[!h]
\centerline{\includegraphics[width=12cm]{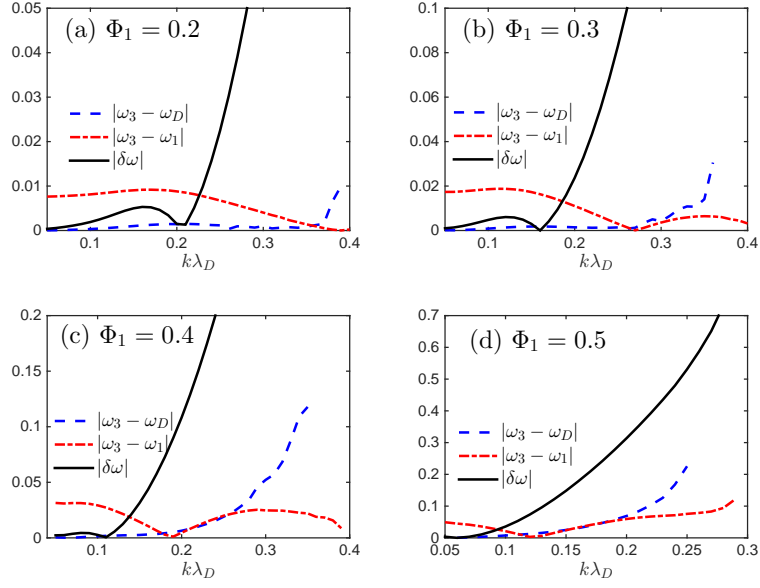}}
\caption{\label{f3} (Color online) Values of $\vert\omega_3-\omega_D\vert$ (blue dashed line), $\vert\omega_3-\omega_1\vert$ (red dashed-dotted line) and $\vert\delta\omega\vert$ (black solid line), normalized to the plasma frequency $\omega_{pe}$,  as a function of $\kld$ for given values of $\Phi_1$. Panel (a) is for $\Phi_1=0.2$, panel (b) is for $\Phi_1=0.3$, panel (c) is for $\Phi_1=0.4$ and panel (d) is for $\Phi_1=0.5$.}
\end{figure}

Figs.~\ref{f3} (a)-(d) show that $\omega_D$ is quite accurate whenever $\Phi_1\alt0.5$ and $\kld\alt0.2$. For the latter range in $\Phi_1$ and $\kld$, the worst accuracy is when $\Phi_1=0.5$ and $\kld=0.2$, $\vert\omega_3-\omega_D\vert\approx\vert\delta\omega\vert/5$, and the accuracy gets rapidly much better when either $\Phi_1$ or $\kld$
 decreases. For example, when $\Phi_1=0.4$ and $\kld=0.2$, $\vert\omega_3-\omega_D\vert\approx\vert\delta\omega\vert/20$. Moreover, $\omega_D$ is more accurate than $\omega_1$ for small values of $\kld$, but within a narrower range in $\kld$ for larger values of $\Phi_1$. For example, Figs.~\ref{f3} (a) and (b) show that $\omega_D$ is more accurate than $\omega_1$ whenever $\kld\alt0.35$ when $\Phi_1=0.2$, but only whenever $\kld\alt0.25$ when $\Phi_1=0.3$. As may be seen in Fig.~\ref{f3} (c), when $\Phi_1=0.4$, $\omega_D$ is more accurate than $\omega_1$ whenever $\kld\alt0.2$ and nearly as accurate as $\omega_1$ when $0.2\alt\kld\alt0.25$.  Fig.~\ref{f3} (d) shows that, when $\Phi_1=0.5$, $\omega_D$ is more accurate than $\omega_1$ whenever $\kld\alt0.15$ and nearly as accurate as $\omega_1$ when $0.15\alt\kld\alt0.2$. Therefore, we conclude that the nonlinear frequency derived using Dawson's potential is more accurate than that calculated with a sinusoidal potential whenever $\kld\alt0.2$ and $\Phi_1\alt0.5$, which supports the conclusions drawn in Paragraph~\ref{II.B.1} by comparing the fields profiles. As for $\omega_1$, Figs.~\ref{f3} (a) and (b) show that $\vert\omega_3-\omega_1\vert$ rapidly decreases compared to $\vert\delta\omega\vert$ when $\kld\agt0.2$, and Figs.~\ref{f3} (b)-(d) show that $\vert\omega_3-\omega_1\vert<\vert\delta\omega\vert$ whenever $\Phi_1\ge0.3$ and $\kld>0.2$. For example, $\vert\omega_3-\omega_1\vert<\vert\delta\omega\vert/7$ when $\kld=0.25$ and $0.3\leq\Phi_1\leq0.5$. When $\Phi_1=0.2$ and $\kld=0.2$, Fig.~\ref{f3} (a) shows that $\vert\omega_3-\omega_1\vert$ is larger than $\vert\delta\omega\vert$, but quickly decreases compared to $\vert\delta\omega\vert$ as $\kld$ increases. Moreover, when $\Phi_1=0.2$, $\vert\omega_3-\omega_1\vert<10^{-2}\omega_{pe}$ whatever $\kld$, so that $\omega_1$ remains very close to $\omega_3$. Therefore, in agreement with the results of Paragraph ~\ref{II.B.1}, we conclude that a harmonic potential yields accurate estimates for the nonlinear frequency whenever $\kld\agt0.2$, although Dawson's potential may yield more accurate results for small amplitudes. Moreover, better results are obtained with a sinusoidal potential than with Dawson's one whenever $\kld\agt0.25$ and $\Phi_1\agt0.3$.
\begin{figure}[!h]
\centerline{\includegraphics[width=12cm]{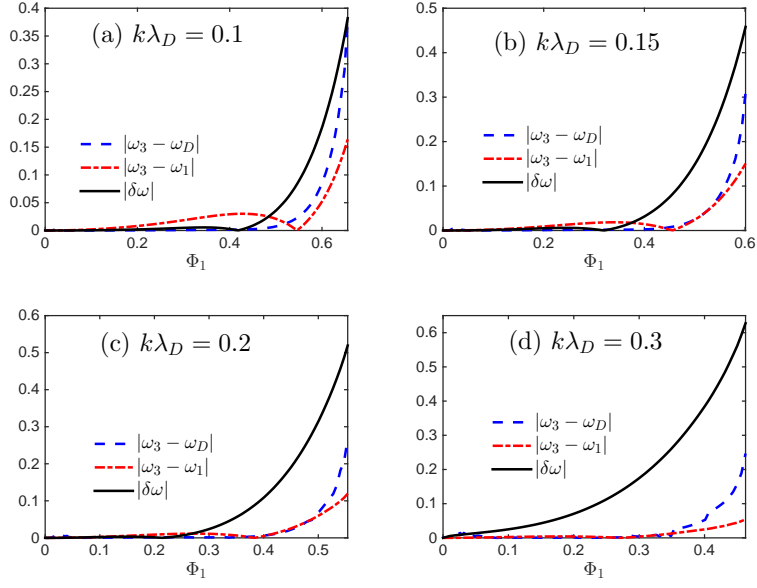}}
\caption{\label{f3b} (Color online) Values of $\vert\omega_3-\omega_D\vert$ (blue dashed line), $\vert\omega_3-\omega_1\vert$ (red dashed-dotted line) and $\vert\delta\omega\vert$ (black solid line), normalized to the plasma frequency $\omega_{pe}$, as a function of $\Phi_1$ for given values of $\kld$. Panel (a) is for $\kld=0.1$, panel (b) is for $\kld=0.15$, panel (c) is for $\kld=0.2$ and panel (d) is for $\kld=0.3$.}
\end{figure}

These conclusions may also be appreciated from Fig.~\ref{f3b} plotting $\vert\omega_3-\omega_D\vert$, $\vert\omega_3-\omega_1\vert$ and $\vert\delta\omega\vert$ as a function of $\Phi_1$, for fixed values of $\kld$.  When $\kld=0.1$, Fig.~\ref{f3b} (a) shows that $\vert\omega_3-\omega_D\vert<\vert\delta\omega\vert/5$ whenever $\Phi_1\alt0.5$ (except close to the region when $\delta \omega$ changes sign), so that $\omega_D$ is quite accurate for this range of amplitudes. However, Fig.~\ref{f3b} (a) also  shows that $\omega_1$ happens to be more accurate than $\omega_D$ when $\Phi_1\agt0.5$. This is quite unexpected because, as may be clearly seen in Fig.~\ref{f1}, the profile of the adiabatic electrostatic field is much closer to Dawson's one than to a sinusoid. The good accuracy of $\omega_1$ is due to the fact that it happens to match $\omega_3$ when $\Phi_1\approx0.55$, which lets it be more accurate than $\omega_D$ for large amplitudes. However, neither $\omega_1$ nor $\omega_D$ are accurate for the largest amplitudes, close to the wave breaking limit. Moreover, although this may not be seen in Fig.~\ref{f3b}, using Dawson's potential allows for solutions to the nonlinear dispersion relation over a narrower range in $\Phi_1$ than when using the adiabatic potential.  Indeed,  when $\kld=0.1$, solutions only exist when $\Phi_1<0.66$ with Dawson's potential, instead of $\Phi_1<0.71$ with the adiabatic one. Hence, Dawson's potential cannot be used for the largest wave amplitudes. This is true whatever $\kld$. For example, one may see in Fig.~\ref{f3} (d) that, when $\Phi_1=0.5$, $\vert\omega_3-\omega_D\vert$ is only plotted up to $\kld=0.25$, unlike $\vert\omega_3-\omega_1\vert$ which is plotted up to $\kld=0.29$. This is because, using Dawson's potential, we could not solve the dispersion relation beyon $\kld=0.25$ when $\Phi_1=0.5$. 

By comparing the results obtained with the four values of $\kld$ considered in Fig.~\ref{f3b} (a)-(d), one clearly sees that the accuracy of $\omega_1$ increases with $\kld$. Fig.~(\ref{f3b}) (d) shows that it is excellent when $\kld=0.3$, $\vert\omega_3-\omega_1\vert<\vert\delta\omega\vert/10$ whatever $\Phi_1$. Fig.~\ref{f3b} (c) shows that it is also very good when $\kld=0.2$, although $\vert\omega_3-\omega_1\vert>\vert\delta\omega\vert$ when $\Phi_1\alt0.25$. However, $\omega_1$ remains very close to $\omega_3$ for such small amplitudes, $\vert\omega_3-\omega_1\vert<10^{-2}\omega_{pe}$. Hence, we conclude again that a harmonic potential yields accurate results for the nonlinear frequency whenever $\kld\agt0.2$. As for Dawson's potential, it yields quite accurate results whenever $\kld\alt0.2$ and $\Phi_1\alt0.5$.

\section{Maximum amplitude for an adiabatic electron plasma wave growing in a Maxwellian plasma}
\label{III}
In~\cp, we already showed that there was no solution to the nonlinear adiabatic dispersion relation beyond a maximum value, $\Phi_1^{\max}$, that depended on $\kld$. Moreover, we noted that $\Phi_1^{\max}$ could only be accurately derived by accounting for the fact that the wave frame was not inertial, which made the nonlinear electron distribution a nonlocal function of the phase velocity. 

In this Section, we discuss in detail the reason why we cannot solve the dispersion relation beyond~$\Phi_1^{\max}$, and what this implies for slowly varying EPW's. 
\begin{figure}[!h]
\centerline{\includegraphics[width=12cm]{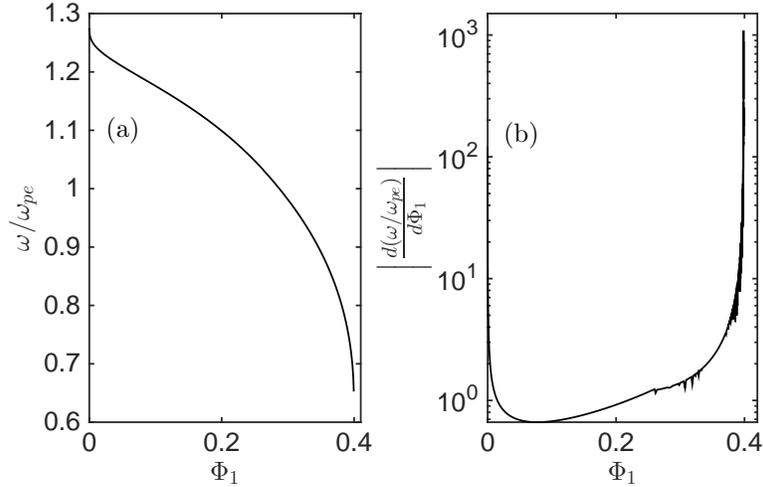}}
\caption{\label{f4} Panel (a), $\omega/\omega_{pe}$ and, panel (b), $\left\vert\frac{d(\omega/\omega_{pe})}{d\Phi_1}\right\vert$, as a function of $\Phi_1$ when $\kld=0.4$. No solution to the dispersion relation could be found when $\Phi_1>\Phi_1^{\max}\approx 0.4$.}
\end{figure}

As may be seen in~Fig.~\ref{f4} when $\kld=0.4$, the values of $\omega$ solving the nonlinear adiabatic dispersion relation seem to be such that $d\omega/d\Phi_1\rightarrow-\infty$ when $\Phi_1\rightarrow \Phi_1^{\max}$. Then, clearly, no solution to the adiabatic dispersion relation can be found when $\Phi_1>\Phi_1^{\max}$.

Now, the adiabatic dispersion relation is only valid when $\vert d\omega/dt\vert$ is small enough. If $\gamma$ is the wave growth rate, the latter condition translates into $(\gamma \Phi_1)\vert d\omega/d\Phi_1\vert$ be small enough. Since $\vert d\omega/d\Phi_1\vert\rightarrow+\infty$ when $\Phi_1\rightarrow \Phi_1^{\max}$, we conclude that there exists a maximum amplitude, $\Phi_1^{\sup}(\gamma)<\Phi_1^{\max}$, beyond which the adiabatic dispersion relation is no longer valid. Moreover, for small enough $\gamma$'s, $\Phi_1^{\sup}(\gamma)\approx \Phi_1^{\sup}(0)=\Phi_1^{\max}$. 

Then, the question remains to know whether there can be any solution to the EPW dispersion relation when $\Phi_1>\Phi_1^{\sup}(\gamma)$. If such a solution existed, the dispersion relation would necessarily be nonadiabatic. Consequently, $\omega$ would decrease very rapidly with $\Phi_1$ whenever $\Phi_1>\Phi_1^{\sup}(\gamma)$, so that $\omega\rightarrow -\infty$ when $\Phi_1\approx \Phi_1^{\max}$, which would be unphysical. Hence, there cannot be any solution to the EPW dispersion relation when $\Phi_1\agt\Phi_1^{\max}$. A nearly monochromatic slowly growing wave cannot exist when $\Phi_1>\Phi_1^{\max}$. The wave necessarily breaks. 
\begin{figure}[!h]
\centerline{\includegraphics[width=12cm]{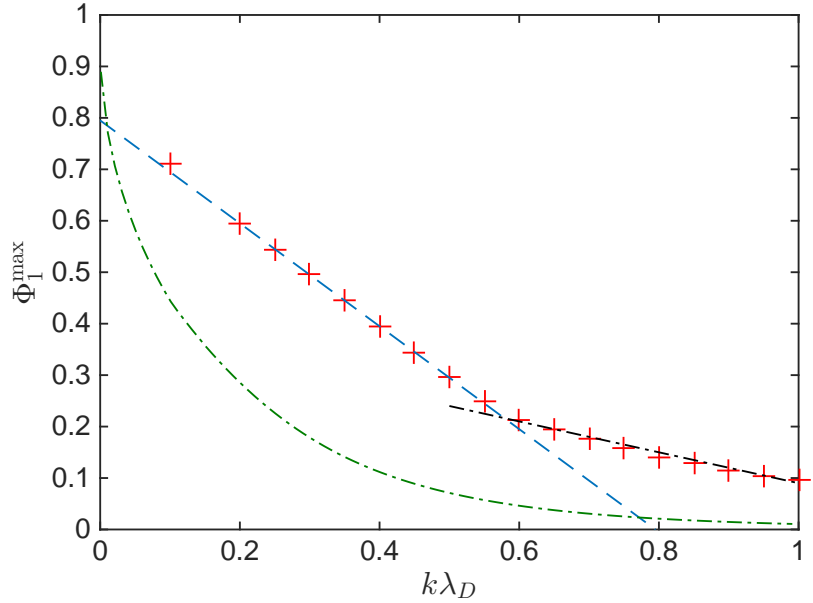}}
\caption{\label{f5} (Color online) The red pluses are the values found for $\Phi_1^{\max}$ from the theory of~\cp, while the green dashed-dotted line plots the values of $\Phi_1^{\max}$ found by Coffey in~\rf{coffey}. The blue dashed line is $\Phi_1^{\max}=0.795-\kld$ while the back dashed-dotted line is $\Phi_1^{\max}=0.39-0.3\kld$.}
\end{figure}

The values found for $\Phi_1^{\max}$ are plotted in Fig.~\ref{f5} as a function of $\kld$, when $0.1<\kld<1$. When $\kld\alt0.6$, $\Phi_1^{\max}$ is very well approximated by $\Phi_1^{\max}\approx 0.795-\kld$, while when $0.6\alt\kld<1$, $\Phi_1^{\max}\approx 0.39-0.3\kld$. 

The values we derive for $\Phi_1^{\max}$ are systematically smaller than the wave breaking limit, $\mathfrak{E}=1$, given by Dawson in~\rf{dawson}. Indeed,  from~Eq.~(\ref{phi1d}), $\mathfrak{E}=1$ corresponds to,
\begin{eqnarray}
\nonumber
\Phi_1^{\text{Dawson}}&=&J_0(1)-J_2(1)+\frac{J_1(1)+J_3(1)}{2}\\
&\approx&0.88.
\end{eqnarray}
This is because Dawson only requires that the electric field has to be single-valued, while we impose the more restrictive condition that a solution to the dispersion relation  must exist. 

 Moreover, Fig.~\ref{f5}~shows that our values for $\Phi_1^{\max}$ are larger than the wave breaking limit derived by Coffey \cite{noteE}~(except, maybe, when $\kld<0.02$).  There are two reasons for such a discrepancy. First, Coffey assumes that the unperturbed distribution function is a waterbag, while we assume that it is a Maxwellian. Second, unlike Coffey, we account for the whole past history of the wave in order to derive $\Phi_1^{\max}$. In particular, the values plotted in Fig.~\ref{f5} are for a wave that has kept on growing in a homogenous plasma. The nonlinear dispersion relation would change if the time variations of the wave amplitude were not monotonous, or if the wave propagated in an inhomogeneous plasma~(see~\rf{benisti17}). Consequently, the values of $\Phi_1^{\max}$ are expected to depend on the particular way the wave has reached the amplitude $\Phi_1^{\max}$. Therefore, it is impossible to derive \textit{a priori}~a wave breaking limit valid in any situation.
 
 Note that we choose to plot the wave breaking limit as a function of the amplitude of the first harmonic of the potential, because this yields very simple scaling laws. However, the wave breaking limit is usually defined as a function of $\delta n_e/n_e$, where $\delta n_e$ is the density fluctuation induced by the wave. From Poisson equation and with our normalization, $\delta n_e/n_e=-\sum_jj^2\Phi_j\cos(j\varphi)$, which significantly departs from the sinusoidal approximation, $-\Phi_1\cos(\varphi)$, for the largest wave amplitudes. In particular, the minimum value of $\delta n_e$ over $\varphi$, which we denote by $\delta n_e^{\min}$, is significantly smaller than $\Phi_1$ while its maximum value,  $\delta n_e^{\max}$, is significantly larger than $\Phi_1$. For example, when $\kld=0.14$ (a situation we investigate in detail in Section~\ref{V}), we find $\Phi_1^{\max}\approx 0.65$, which corresponds to $\delta n_e^{\max}/n_e\approx 1.04$ and $-\delta n_e^{\min}/n_e\approx 0.43$.  The latter estimate for $-\delta n_e^{\min}/n_e$ is in good agreement with the minimum density derived in the PIC simulation of Ref.~\onlinecite{rousseaux} just before the wave starts to break. Indeed, as illustrated in Fig.~\ref{phi_surf} (d), $-\delta n_e^{\min}/n_e\alt0.4$ just before wave breaking in the PIC simulation, which would correspond to $\Phi_1^{\max}\alt0.59$. Hence, at least for the example studied in Section~\ref{V}, we could check that values plotted for  $\Phi_1^{\max}$ in Fig.~\ref{f5} do  yield an upper bound for the wave breaking limit, which is close to the actual limit. 
 
In general, an EPW breaks because of the unstable growth of secondary modes due, for example, to the trapped particle instability~\cite{sudan,brunner04,brunner,friou}. This has been clearly shown in~\rf{friou} for an SRS-driven EPW.  Wave breaking occurs when the secondary modes grow so fast that their amplitude eventually overtakes that of the EPW. Accurately describing such a complex situation is a difficult task, which is part of our research program. However, regardless of the reason why the EPW should break, the values plotted for $\Phi_1^{\max}$ in Fig.~\ref{f5} do provide a rigorous upper bound for the amplitude of an adiabatic wave growing in a uniform plasma. To the best of our knowledge, such a rigorous result was not available in previous publications. Moreover, the theory of~\cp~is general enough to address any situation, regardless of the time and space evolution of the wave amplitude and plasma density. Therefore, the procedure described in this Section may be applied to derive the wave breaking limit in any physics situation, provided that the EPW varies slowly enough.

\section{Envelope equation for adiabatic electron plasma waves}
\label{IV}
One of the most important issues, regarding nonlinear EPW's, is the ability to predict their space and time variations. Envelope equations have proven to be a very effective and accurate way to do so, as shown in Refs.~\onlinecite{brama,benisti10,benisti12,benisti18}~for an EPW driven by SRS in an initially uniform Maxwellian plasma. Moreover, envelope equations valid in a nonstationary and non-uniform situation have been derived in Refs.~\onlinecite{dodin1}~and~\onlinecite{benisti16} by resorting to a variational formalism. However, in the latter articles, the envelope equations have been written in a rather formal way, where the role played by the vector potential did not appear clearly, nor did the space-dependence of the scalar potential. In this Section, we provide explicit expressions for the nonlinear envelope equation of a driven EPW, valid whatever $\kld$ and up to amplitudes close to the wave breaking limit.
\subsection{General results}
\label{IV.1}
The Lagrangian density for the self-consistent wave-particle interaction, as derived in Refs.~\onlinecite{dodin1}~and~\onlinecite{benisti16},  reads
\begin{equation}
\mL=\varepsilon_0\frac{k^2\phi_A^2}{4}-\frac{(\nabla\times A_0)^2}{2\mu_0}-\mL_u-\mL_t,
\end{equation}
where $A_0$ is the vector potential, and where $(k^2\phi_A^2)/2$ is the averaged value of the electrostatic field squared. Namely, using the same notation as in \cp, the electrostatic field reads $E_{el}=\sum_{n\geq 1}E_n\sin(n\varphi)$. Then, $(k\phi_A)^2=\sum_{n\geq 1}E_n^2$. Moreover, since we only look for an envelope equation at first order in the space and time derivatives of the fields, it is enough to derive the electrostatic potential, $\phi$, at zeroth order. Hence, it may be approximated by $\phi=\sum_{n\geq 1}\phi_n\cos(n\varphi)$, with $\phi_n\approx E_n/nk$. Then, 
\begin{equation}
\label{newnew}
\phi_A^2= \sum_{n\geq 1}n^2\phi_n^2.
\end{equation}
 As for $\mL_u$ and $\mL_t$, they read
\begin{eqnarray}
\label{Lu}
\mL_u&=&\int_{\vert P-mV_\phi\vert>m\mA_s}f(X,P,t)\mH_udP,\\
\label{Lt}
\mL_t&=&\int_{0}^{m\mA_s}f(I,\bm{x})\mH_td(kI),
\end{eqnarray}
where
\begin{eqnarray}
\mH_u&=&H+Pv_\phi-\frac{mV_\phi^2}{2}+\frac{e^2A_0^2}{2m},\\
\mH_t&=&H-\frac{mV_\phi^2}{2}+\frac{e^2A_0^2}{2m},
\end{eqnarray}
with 
\begin{eqnarray}
v_\phi&=&\omega/k,\\
\label{Vphi}
V_\phi&=&\omega/k-eA_0/m,\\
H&=&\frac{(kp-mV_\phi)^2}{2m}-e\phi,
\end{eqnarray}
where $kp=mv-eA_0$, $v$ being the electron velocity. Note that $H$ is $m$ times the Hamiltonian defined in~\cp, so as to make it scale as an energy. Moreover, in Eq.~(\ref{Lt}) for $\mL_t$, $I$ is the action for the Hamiltonian $H$, while in Eq.~(\ref{Lu}) for $\mL_u$,  $P=kI$, and $X$ is canonically conjugated to $P$ for $\mH_u$.  In Eqs.~(\ref{Lu})~and~(\ref{Lt}), $\mA_s$ is defined like in~\cp, $4\pi \mA_s$ is the width (in velocity) of the separatrix. As for $f$, it is the adiabatic electron distribution function. It is normalized so that $\int fd(kI)=n_e$, where $n_e$ is the electron density. $f$ is $(n_e/m)$ times the function derived in~\cp, because $kI$ now scales as a momentum and not as a velocity. 

As discussed in~\cp, one may resort to the adiabatic approximation to derive the envelope equation of a growing EPW if, for all electrons, 
\begin{equation}
\label{4cp}
\gamma T_B\alt 1/2,
\end{equation}
where $\gamma$ is the wave growth rate, as seen by the electron, and $T_B$ is the period of a deeply trapped orbit. Then, the envelope equation for a driven wave reads~\cite{benisti16}
\begin{eqnarray}
\nonumber
\frac{\varepsilon_0E_1E_d}{2}\cos(\delta\varphi_d)&=&\partial_{t\omega}\mL\vert_{\mA_s}+\bm{\nabla}.\partial_{\bm{k}}\mL\vert_{\mA_s}\\
\label{enva}
&&-\int_0^{\frac{m\mA_s}{k}} \mH_t\frac{\bm{k}}{k}.\bm{\nabla}f(I,\bm{x})dI,
\end{eqnarray}
where $E_d$ is the drive amplitude (assumed to be sinusoidal) and $\delta \varphi_d$ is the phase difference between the drive and the electrostatic field. Moreover, the symbol $\vert_{\mA_s}$ means that the integral boundaries in Eq.~(\ref{Lu})~and~(\ref{Lt}) are not to be derived or, more precisely, that the fractions of trapped and untrapped electrons are to be considered as constants. Namely,
\begin{eqnarray}
\nonumber
\partial_{t\omega}\mL_u\vert_{\mA_s}&=&\int_{\vert P-mV_\phi\vert>m\mA_s}\partial_t\left[f(X,P,t)\partial_\omega\mH_u\right]dP,\\
\nonumber
\partial_{t\omega}\mL_t\vert_{\mA_s}&=&\int_{0}^{m\mA_s}f(I,\bm{x})\partial_{t\omega}\mH_td(kI),\\
\nonumber
\bm{\nabla}.\partial_{\mk}\mL_u\vert_{\mA_s}&=&\int_{\vert P-mV_\phi\vert>m\mA_s}\bm{\nabla}.\left[f(P,X,t)\partial_{\mk}\mH_u\right]dP,\\
\nonumber
\bm{\nabla}.\partial_{\mk}\mL_t\vert_{\mA_s}&=&\int_0^{\frac{m\mA_s}{k}}\bm{\nabla}.\left\{f(I,\bm{x})\left[\frac{\bm{k}}{k}\mH_t-k\partial_{\bm{k}}\mH_t\right]\right\}dI.
\nonumber
\end{eqnarray}
Moreover,
\begin{eqnarray}
\label{coeff1}
\partial_\omega\mH_u&=&\partial_\omega H+\frac{P}{k}-\frac{mV_\phi}{k},\\
\partial_{\bm{k}}\mH_u&=&\partial_{\bm{k}}H-\frac{P\omega\bm{k}}{k^3}+\frac{mV_\phi\omega\mk}{k^3},\\
\partial_\omega\mH_t&=&\partial_\omega H-\frac{mV_\phi}{k},\\
\label{coeffn}
\partial_{\bm{k}}\mH_t&=&\partial_{\bm{k}}H+\frac{mV_\phi\omega\bm{k}}{k^3}.
\end{eqnarray}
Note that the vector potential explicitly enters the envelope equation through $V_\phi=\omega/k-eA_0/m$. In addition to $A_0$ which follows from Eqs.~(32) of~\cp, one only needs to derive $\partial_\omega H$ and $\partial_{\mk} H$ in order to find an explicit expression for the EPW envelope equation. 

For untrapped electrons, 
\begin{equation}
\label{P}
P=\frac{1}{2\pi}\oint\sqrt{2m(H+e\phi)}d\varphi+\eta mV_\Phi,
\end{equation}
where $\eta=+1$ for orbits above the separatrix and $\eta=-1$ below the separatrix, so that
\begin{eqnarray}
\partial_{\omega}H&=&-\eta \frac{m\Omega}{k^2},\\
\partial_{\mk}H &=&\eta \frac{m\omega\Omega}{k^4}\bm{k}=-\frac{\omega\mk}{k^2}\partial_\omega H,
\end{eqnarray}
where 
\begin{eqnarray}
\Omega&=&k\partial_PH\\
&=& \frac{2\pi k}{\sqrt{m}}\left[\oint\frac{d\varphi}{\sqrt{2(H+e\phi)}}\right]^{-1}.
\end{eqnarray}

For trapped electrons, 
\begin{equation}
\label{28}
kI=\frac{1}{4\pi}\oint\sqrt{2m(H+e\phi)}d\varphi,
\end{equation}
so that
\begin{eqnarray}
\partial_{\bm{k}}H &=&\frac{\Omega I}{k^2}\bm{k},\\
\partial_{\omega}H&=0,
\end{eqnarray}
where 
\begin{eqnarray}
\nonumber
\Omega&=&\partial_IH\\
\label{Omega}
&=& \frac{4\pi k}{\sqrt{m}}\left[\oint\frac{d\varphi}{\sqrt{2(H+e\phi)}}\right]^{-1}.
\end{eqnarray}
Hence, the explicit expression of the nonlinear EPW envelope equation follows from the sole derivation of $\Omega$, which may only be performed once the $\varphi$-variations of $\phi$ are known. 

Nevertheless, simple approximations for $\Omega$ are easily obtained. For untrapped electrons whose orbits are far away from the separatrix, 
\begin{equation}
\label{34}
\Omega \approx k(P/m-\eta V_\phi).
\end{equation}
For trapped orbits far away from the separatrix, 
\begin{equation}
\label{35}
\Omega \approx 2k\sqrt{\frac{e\phi''(0)}{m}},
\end{equation}
where $\phi''(0)\equiv d^2\phi/d\varphi^2$ calculated at the $O$-point.

Moreover, let $\phi_m$ be the minimum of $\phi$ over one wavelength, assumed to be reached at the $X$-point (which always happens for the situations considered in~\cp). Then, for orbits very close to the separatrix, $\Omega$ goes to zero as
\begin{equation}
\label{marre}
\Omega \sim \frac{\pi k\sqrt{\vert\phi''(\pi)\vert}}{2\sqrt{m}\ln(2\pi/\varepsilon)},
\end{equation}
where $\phi''(\pi)\equiv d^2\phi/d\varphi^2$ calculated at the $X$-point, and where
\begin{equation}
\varepsilon=\left\vert\frac{H+\phi_m}{\phi_m}\right\vert.
\end{equation}

\subsection{Sinusoidal potential}
As discussed in Section~\ref{II}, a sinusoidal approximation is accurate for a growing wave, whatever the wave amplitude, provided that $\kld\agt0.2$. Then,  for a sinusoidal potential, $\Omega$ is given by the following formulas.

For untrapped electrons,
\begin{equation}
\Omega=\frac{\pi\omega_B}{\sqrt{\zeta_u}K_1(\zeta_u)},
\end{equation}
where $K_1$ is the elliptic integral of first kind~\cite{abramowitz}, where, 
\begin{equation}
\omega_B=k\sqrt{\frac{e\phi_1}{m}},
\end{equation}
is the so called bounce frequency, and where $\zeta_u$ is related to $P$ by
\begin{equation}
P-\eta mV_\phi=\frac{4}{\pi}\sqrt{me\phi_1}\frac{K_2(\zeta_u)}{\sqrt{\zeta_u}},
\end{equation}
where $K_2$ is the elliptic integral of second kind~\cite{abramowitz}. When $\zeta_u<0.85$, $\Omega$ differs from the approximate expression, Eq.~(\ref{34}), by less than 10\%.

For trapped electrons, 
\begin{equation}
\Omega=\frac{\pi\omega_B}{K(\zeta_t)},
\end{equation}
where $\zeta_t$ is related to $I$ by
\begin{equation}
kI=\frac{4}{\pi}\sqrt{me\phi_1}\left[K_2(\zeta_t)+(\zeta_t-1)K_1(\zeta_t)\right].
\end{equation}
When $\zeta_t<0.6$, $\Omega$ differs from the approximate expression, Eq.~(\ref{35}), by less than 25\%. 

\subsection{Dawson's potential}
As discussed in Section~\ref{II}, the field profile proposed by Dawson is very close to the adiabatic one whenever $\kld\alt0.2$, and up to values close to the wave breaking limit. Moreover, the nonlinear frequency derived from Dawson's potential, $\omega_D$, was shown in Section~\ref{II} to be quite accurate whenever $\kld\alt0.2$ and $\Phi_1\alt 0.5$. As regards the envelope equation derived using Dawson's potential, it is expected to be accurate up to amplitudes close to the wave breaking limit. Indeed, as discussed in Paragraph~\ref{IV.1}, the coefficients of this equation mainly depend on $\Omega$ and, from Eqs.~({\ref{coeff1})-(\ref{coeffn}), on $\omega$ and $V_\phi$. In Section~\ref{II}, we found that replacing $\omega$ by $\omega_D$ would entail an error much less than the nonlinear frequency shift, $\delta \omega$, only when $\kld\alt0.2$ and $\Phi_1\alt 0.5$. However, unless the wave amplitude is close to the wave breaking limit, $\delta \omega \ll \omega$, and it is valid to replace $\omega$ with $\omega_D$ in Eqs.~(\ref{coeff1})-(\ref{coeffn}). The same conclusion holds for the value of $\Omega$ calculated for passing particles away from the separatrix, and whose expression is given by Eq.~(\ref{34}). As for trapped particles away from the separatrix,  Eq.~(\ref{35}) shows that $\Omega$ is proportional to $\sqrt{\phi''(0)}$, which is always very well estimated using Dawson's potential, even for the largest amplitudes. For example, when $\kld=0.1$ and $\mathcal{E}_{\max}\approx 0.745$, which corresponds to Fig.~\ref{f1} (a) of Section~\ref{II}, $\sqrt{\phi''(0)}$ is only underestimated by 8\% when using Dawson's potential (it would be overestimated by 22\% with a sinusoidal potential). $\Omega$, for particles close to the separatrix, would not be correctly calculated with Dawson's potential. Nevertheless, the corresponding values as given by Eq.~(\ref{marre}) are small, leading to a small contribution to the envelope equation, so that their accurate estimate is not essential. Hence, we conclude that the envelope equation derived using Dawson's potential should be accurate whenever $\kld\alt0.2$ and up to amplitudes close to the wave breaking limit. This may be appreciated in Fig.~\ref{fvg} plotting the group velocity, $v_g$, as a function of $\Phi_A$ when $\kld=0.14$.

Using Eq.~(\ref{phi}) one finds that, with Dawson's potential, $\Omega$ is given by the following formulas.

From Eq.~(\ref{Omega}), $\Omega$ for untrapped electrons is,
\begin{equation}
\frac{\pi\omega_{pe}}{\Omega}=\int_0^{\pi} \frac{\left[1+\mathfrak{E}\cos(\varphi_0)\right]d\varphi_0}{\sqrt{2[h+\Phi(\varphi_0)]}},
\end{equation}
where $h=k^2H/m\omega_{pe}^2$ and where $\Phi(\varphi_0)$ is given by Eq.~(\ref{9}). Moreover, the relation between $h$ and $P$ follows from Eq.~(\ref{P}), which reads,
\begin{equation}
P-\eta mV_\phi=\frac{m\omega_{pe}}{k\pi}\int_{0}^{\pi}\sqrt{2[h+\Phi(\varphi_0)]}[1+\mathfrak{E}\cos(\varphi_0)]d\varphi_0.
\end{equation}
For trapped electrons, 
\begin{equation}
\frac{\pi\omega_{pe}}{\Omega}=\int_0^{\varphi_{\max}} \frac{\left[1+\mathfrak{E}\cos(\varphi_0)\right]d\varphi_0}{\sqrt{2[h+\Phi(\varphi_0)]}},
\end{equation}
with $h+\Phi(\varphi_{\max})=0$. Moreover, from Eq.~(\ref{28}), $h$ is related to $I$ by ,
\begin{equation}
kI=\frac{m\omega_{pe}}{k\pi}\int_{0}^{\varphi_{\max}}\sqrt{2[h+\Phi(\varphi_0)]}[1+\mathfrak{E}\cos(\varphi_0)]d\varphi_0.
\end{equation}

\subsection{Envelope equations valid whatever the wave amplitude}
The envelope equation, Eq.~(\ref{enva}), is only valid when the wave amplitude is so large that all electrons may be considered adiabatic. However, in general, the condition for adiabaticity, as given by Eq.~(\ref{4cp}), is only fulfilled by a fraction, $\mF_a$, of electrons. Moreover, as shown  in Refs.~\onlinecite{benisti16}~and~\onlinecite{benisti07}, the contribution to the envelope equation from most non-adiabatic electrons  is the linear one,
\begin{equation}
\label{lin}
\partial_{t\omega} \left(\chi E_1^2\right)-\bm{\nabla}\partial_{\mk} \left(\chi E_1^2\right)+2\nu_L\partial_\omega\chi E_1^2=\frac{\varepsilon_0E_1E_d}{2}\cos(\delta\varphi_d),
\end{equation}
where $\nu_L$ is the Landau damping rate, 
\begin{equation}
\label{lineaire}
\nu_L \equiv -\frac{\pi e^2}{\varepsilon_0m k^2\partial_\omega\chi}f'_0(x,v_\phi,t),
\end{equation}
$f'_0$ being the derivative, with respect to velocity, of the unperturbed velocity distribution function, and $\chi$ being the adiabatic limit of the linear electron susceptibility,
\begin{equation}
\chi =-\frac{e^2}{\varepsilon_0mk} P.P.\left(\int\frac{f'_0}{kv_0-\omega}dv_0\right).
\end{equation}
Then, as shown in~Refs.~\onlinecite{brama,benisti10,benisti12,benisti18,benisti16,benisti07,benisti09}, the envelope equation of a slowly varying wave, valid whatever its amplitude, is the sum of the adiabatic envelope equation, Eq.~(\ref{enva}), multiplied by $\mF_a$, and of the linear envelope equation, Eq.~(\ref{lin}), multiplied by $(1-\mF_a)$. Moreover, by using the results derived for a sinusoidal potential when $\kld\geq0.2$, one obtains an explicit expression for the envelope equation of a growing wave, valid whatever its amplitude, without having to derive the potential self-consistently as in~\cp. When $\kld<0.2$, the results derived using Dawson's potential also provide explicit expressions for the envelope equation, but only up to an amplitude close to the wave breaking limit. Furthermore, as discussed in Section~\ref{II}, the envelope equation derived by assuming a growing wave remains valid even when the wave amplitude does not keep growing, provided that $V_\phi$ varies less rapidly than $\mA_s$.

\subsection{Approximate expression for the envelope equation}
\label{IV.5}
In this Section, we specialize to plasmas which are essentially uniform, so that the last term in the right-hand side of Eq.~(\ref{enva}) is negligible. Moreover, the Lagrange equation, $\partial_{\phi_A}\mL=0$, reads $1+\chi_a=0$, where,
\begin{eqnarray}
\nonumber
\chi_a=-\frac{2}{\varepsilon_0k^2\phi_A}\left[\int_{\vert P-mV_\phi\vert>m\mA_s}f(X,P,t)\partial_{\phi_A}H dP\right.\\
\label {chia}
+\left.\int_0^{m\mA_s}f(I,\bm{x})\partial_{\phi_A}Hd(kI)\right].
\end{eqnarray}
Now, it is easily shown that,
\begin{eqnarray}
\mH_u&=&\int_0^{\phi_A}\frac{\partial H}{\partial {\phi'_A}}d\phi'_A,\\
\mH_t&=&\int_0^{\phi_A}\frac{\partial H}{\partial \phi'_A}d\phi'_A+\frac{mV_\phi^2}{2},
\end{eqnarray}
which lets us write
\begin{equation}
\label{mL2}
\mL=\frac{\varepsilon_0k^2}{2}\int_0^{\phi_A}[1+\chi_a(\phi'_A)]\phi'_Ad\phi'_A+\frac{m\sigma k V_\phi^2}{2},
\end{equation}
where
\begin{equation}
\sigma\equiv \int_0^{\frac{m\mA_s}{k}}f(I,\bm{x})dI.
\end{equation}
From the results of the companion paper~\cp, we know that, except when $\phi_A$ is close the wave breaking limit, the harmonic content of the scalar potential, and the wave frequency, do not vary much with $\phi_A$. This implies that $\chi_a$ does not depend much on the wave amplitude so that, in the integral of Eq.~(\ref{mL2}), one may replace $1+\chi_a(\phi'_A)$ with $1+\chi_a(\phi_A)$. Then,
\begin{eqnarray}
\partial_{\bm{k}}\mL&\approx&-\bm{v}_g\Lambda_a,\\
\partial_\omega\mL&\approx&\Lambda_a+m\sigma V_\phi\left[1-\frac{V_\phi-2v_\phi}{2v_g}\right],
\end{eqnarray}
where,
\begin{equation}
\label{vg}
\bm{v}_g=-\partial_{\mk}\chi_a/\partial_\omega\chi_a,
\end{equation}
and,
\begin{equation}
\label{lambda}
\Lambda_a=\frac{\varepsilon_0 E_A^2}{4}\partial_{\omega}\chi_a-\frac{m\sigma V_\phi}{2v_g}(V_\phi-2v_\phi),
\end{equation}
\begin{figure}[!h]
\centerline{\includegraphics[width=8.6cm]{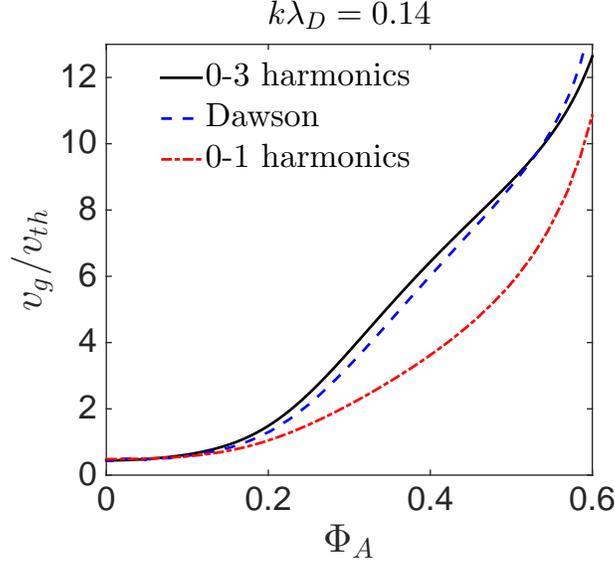}}
\caption{\label{fvg} (Color online) Variations of $v_g$ as a function of $\Phi_A=(k\lambda_D)^2\phi_A$, when $k\lambda_D=0.14$. The black solid line is for the self-consistent electric field derived from the theory of~\cp~by accounting for harmonics 0 to 3, while the red dashed-dotted line is obtained by only accounting for the zeroth and first harmonics. The blue dashed line represents the values $v_g$ derived by using Dawson's potential, Eq.~(\ref{9}).}
\end{figure}
where we have denoted $E_A=k\phi_A$. $E_A$ may be viewed as the effective amplitude of the electrostatic field (for a sinusoidal wave, $E_A=E_1$, is the field amplitude). As for $v_g$, its variations with $\Phi_A\equiv(k\lambda_D)^2\phi_A$ are illustrated in Fig.~\ref{fvg} when $k\lambda_D=0.14$, for the self-consistant potential derived as in ~\cp, for a purely sinusoidal potential, and for Dawson's potential~\cite{noteD}. It is noteworthy that the nonlinear values of $v_g$ may be significantly (up to 30 times) larger than its linear limit. One may also see in Fig.~\ref{fvg} that the values of $v_g$, derived with the self-consistent potential, are very close to those obtained with Dawson's potential. This shows the relevance of using the latter simple potential to derive the EPW envelope equation. 

Note that, strictly speaking, $v_g$ is only useful to derive the ray equations, i.e., the transport of the eikonal, not the evolution of the wave action. However, in most situations, the time derivative of the last term in the right-hand side of Eq.~(\ref{lambda}) is negligible in the envelope equation, either because $\sigma$ is small, or because the time variations of $V_\phi$, $v_\phi$ and $v_g$ are slow. In such situations, and when all electrons are adiabatic ($\mF_a=1$), the envelope equation for a driven EPW in a homogeneous plasma reads
\begin{equation}
\label{envb}
\partial_t\Lambda_a+\bm{\nabla}.(\bm{v}_g\Lambda_a)=\frac{\varepsilon_0E_dE_1\cos(\delta\varphi_d)}{2}.
\end{equation}
As further discussed in Section~{\ref{V}, using Eq.~(\ref{envb}) greatly simplifies the derivation of the space and time variations of the plasma wave, because the same $\bm{v}_g$ is used in the equation for $\Lambda_a$ and for the ray tracing. Then, it is clear that Eq.~(\ref{envb}) is valid when such effects as the group velocity splitting~\cite{whitham,dodin3} are negligible. 

Moreover, in a uniform plasma, the space and time variations of $\Lambda_a$ are mainly due to those of the wave amplitude, so that Eq.~(\ref{envb}) may be further simplified by assuming
\begin{equation}
\label{58}
\Lambda_a\approx \frac{\varepsilon_0 E_A^2}{4}\partial_{\omega}\chi_a.
\end{equation}
The approximation, Eq.~(\ref{58}), is explicitly used in the simple model introduced in Section~\ref{V} in order to derive the transverse modes resulting from wavefront bowing. 

\section{Transverse modes resulting from wavefront bowing}
\label{V}

As shown in several papers \cite{rousseaux,masson,berger,silva}, when an EPW grows and enters the strongly nonlinear regime where kinetic effects are important, its spectrum enriches in transverse wavenumbers. Moreover, as discussed in the Introduction, there may be two different reasons for the growth of transverse modes. They may result from instabilities, either electrostatic~\cite{berger} or electromagnetic~\cite{masson,silva}, due to trapped particles.  They may also be the consequence of the transverse inhomogeneity in the EPW nonlinear frequency shift, $\delta \omega$, which entails the wavefront bowing~\cite{rousseaux,masson,berger,silva,yin07,yin08,yin,ban11,yin12,yin13}. It is usually impossible to disentangle the role of each effect directly from experimental results and 2-D PIC simulations, as discussed in the detailed analysis of \rf{rousseaux}. In this Section, we estimate the transverse wavenumbers which only result from wavefront bowing. This lets us conclude about the ability to correctly describe the EPW spectrum by only accounting for the latter effect. 

Moreover, SRS essentially occurs where the EPW is nearly monochromatic. Indeed, once secondary modes have grown unstable, the magnitude of the density fluctuations dramatically drops (see~\rf{rousseaux}), and one would expect Thomson scattering rather than Raman scattering. Therefore, the opening angle of the SRS backscattered light is expected to directly follow from the EPW wavefront bowing. This further vindicates the introduction of an accurate and effective model to quantify it. 

\subsection{The ray-in-cell method}
\label{V.1}
In order to address the EPW wavefront bowing, we cannot rely on the numerical methods based on the paraxial~\cite{lax,riazuelo}, nor on the quasioptical~\cite{permitin,dodin_ray}, approximations. Indeed, these do not accurately estimate the transverse variations of the wavenumbers, which are assumed to be small, while we precisely need to derive these variations in order to properly describe wavefront bowing. Consequently, we introduce in this Section the prototype of a new numerical method which we dubbed ray-in-cell (RIC). It combines the resolution of nonstationary ray tracing and envelope equations. The number of quanta for each wave is derived along the rays from the envelope equations. The ray dynamics, from which follow the wavenumbers, is derived from the dispersion relations. For the EPW, the dispersion relation is nonlinear so that the ray dynamics keeps changing 
while the wave is growing. This explains why the ray tracing has to be nonstationary and has to be solved together with the envelope equation. To do so, we first estimate the wave amplitude on a fixed mesh, from an averaging of the wave quanta derived along the rays. This lets us derive the gradient of the wave amplitude on the mesh midpoints, which we project back onto the rays to derive their dynamics (see Paragraph~\ref{V.3} and Fig.~\ref{RIC}).

Actually, the RIC method may be generalized, as in \rf{deb19}, to address multiple wave-wave interaction in various contexts. Indeed, the amplitude of any wave may be estimated on the mesh and then projected onto the rays of any other wave to account for their coupling.  Moreover, solving envelope equations like Eq.~(\ref{enva}) also allows wave-particle interaction (i.e., nonlinear kinetic effects) to be captured. Hence, we expect the RIC method to let us address laser-plasma interaction over space and time scales relevant to inertial confinement fusion (ICF), which is still far from being attainable with kinetic codes. Our long-term objective is to provide quick methods, which can be implemented in the hydrodynamical codes used in ICF, to correctly model laser propagation inside a fusion plasma.

\subsection{A simplified theoretical model}
\label{V.2}
Our modeling of the EPW wavefront bowing rests on several simplifying hypotheses. First, we use the geometrical optics limit, so that the transverse wavenumbers result from the following ray equations,
\begin{eqnarray}
\label{ray1}
d_t\bm{x}_R&=&\partial_{\mk}\Omega_R\vert_{\bm{x},t}, \\
\label{ray2}
d_t\mk_R&=&-\partial_{\bm{x}}\Omega_R\vert_{\mk,t},
\end{eqnarray}
where $\mk_R(t)\equiv\mk[\bm{x}_R(t),t]$, and $\Omega_R[\bm{x},\mk(\bm{x},t),t]\equiv\omega(\bm{x},t)$ solves the EPW nonlinear dispersion relation, $1+\chi_a=0$. Hence, $\partial_{\mk}\Omega_R=\bm{v}_g$ as defined by Eq.~(\ref{vg}). 

Now, the geometrical optics limit usually remains valid as long as most of the EPW energy is not confined within a volume less than $k^{-3}$.  Therefore, it should not be suited to address the EPW self-focussing resulting from wavefront bowing. However, as discussed in Paragraph~\ref{V.3}, the RIC method allows to alleviate most difficulties entailed by self-focussing or ray crossing. This is mainly due to the fact that the wave amplitude is bounded from above, because it is averaged over a grid cell.
 
Moreover, in order to derive $\Omega_R$ in Eqs.~(\ref{ray1}) and (\ref{ray2}), we assume that the longitudinal component of $\mk$ does not change much, and remains much larger than its transverse components. This hypothesis is consistent with the neglect of the $\mk$-rotation, and of the variations of $k\equiv\vert \mk\vert$, to derive $\Omega_R$. Therefore,  $\Omega_R$ may be directly obtained from the results of~\cp, as plotted in Fig.~\ref{f3}. We also restrict to uniform plasmas, so that the space-dependence of $\Omega_R$ directly follows from that of $\Phi_A$, which lets Eq.~(\ref{ray2}) read
\begin{equation}
\label{ray3}
\frac{d\mk_R}{dt}=-\frac{d\Omega_R}{d\Phi_A}\bm{\nabla}\Phi_A.
\end{equation}

Clearly, from Eq.~(\ref{ray3}), the ray equations have to be solved together with the envelope equation for $\Phi_A$. Moreover, the growth of transverse wavenumbers is intrinsically a nonstationary problem, and will be considered as such. Namely, the nonstationarity in $\Omega_R$ follows from the nonstationarity of $\Phi_A$.  

When the EPW results from SRS, Eq.~(\ref{enva}) has to be solved together with the envelope equations for the laser and scattered lights. Solving these coupled equations is a difficult task, which is part of our current research work, but which is way beyond the scope of this paper. Here, we do not aim at an accurate description of SRS. Instead, we want to solve a much simpler problem which is the estimate of the transverse extent of the EPW spectrum. It mainly depends on the growth rate of the transverse wavenumbers, compared to the time it takes for the EPW amplitude to reach the wave breaking limit. Indeed, if the EPW grows very quickly and breaks by the time the $\mk$ direction could change, wavefront bowing is insignificant. By contrast, if the EPW grows very slowly and the laser duration is large enough, significant transverse components in $\mk$ have the time to build up. Hence, in order to correctly estimate the transverse extent of the EPW spectrum, we only need the correct order of magnitude for the EPW growth rate. It is well-known~\cite{benisti18,kruer} that the growth rate of an essentially undamped SRS-driven plasma wave is of the order of,\begin{equation}
\label{gamma0}
\gamma_0 ({\bm x},t) = \frac{e k \, E_{\rm{las}}  ({\bm x},t)}{2 m  \omega_{\rm{las}} \sqrt{ 2 \omega_{s} \partial_{\omega} {\chi_a} }},
\end{equation}
where $\omega_{\rm{las}}$ and $\omega_s$ are, respectively, the laser and scattered wave frequencies ($\omega_s=\omega_{\rm{las}}-\omega$), and where $E_{\rm{las}}$ is the amplitude of the laser electric field. Consequently, 
following the lines of Paragraph~\ref{IV.5}, we use the following simplified envelope equation for the EPW,
\begin{equation}
\label{envc}
\partial_t\Lambda_a+\bm{\nabla}.(\bm{v}_g\Lambda_a)=2\gamma_0\Lambda_a,
\end{equation}
where $\Lambda_a$ is defined by Eq.~(\ref{58}). Let us now introduce $J$ such that $\partial_tJ+\bm{v}_g.\bm{\nabla}J=J\bm{\nabla}.\bm{v}_g$. From Liouville theorem, $J$ is the Jacobian, $J=\vert d\bm{x}_R(t)/d\bm{x}_R(0)\vert$. Then, Eq.~(\ref{envc}) reads
\begin{equation}
\label{envd}
d_tN_p=2\gamma_0N_p,
\end{equation}
where $N_p(t)=J\Lambda_a[\bm{x}_R(t),t]$. Using Eq.~(\ref{envc}) instead of Eq.~(\ref{enva}) greatly simplifies the problem because $\Lambda_a$ and the eikonal have the same characteristics. This allows to easily calculate $N_p$ along a ray. Note that $N_pd\bm{x}_R(0)$ is just the number of plasmons initially located within the infinitesimal volume $d\bm{x}_R(0)$, and calculated along an infinitesimal ray bundle. Then, when $\gamma_0=0$, Eq.~(\ref{envd}) just translates the conservation of the number of plasmons. 

Since we only solve for the EPW amplitude, we cannot account for pump depletion. Then, $E_{\rm{las}}$ in Eq.~(\ref{gamma0}) is given by
\begin{equation}
E_{\rm{las}}=\sqrt{2I_{\rm{las}}/\varepsilon_0 v_{g_{\rm{las}}} },
\end{equation}
where the laser intensity, $I_{\rm{las}}$, is assumed to remain undepleted, and where $v_{g_{\rm{las}}}$ is the laser group velocity.

\subsection{The ray-in-cell numerical scheme}
\label{V.3}
\begin{figure}[!h]
\centerline{\includegraphics[width=8.6cm]{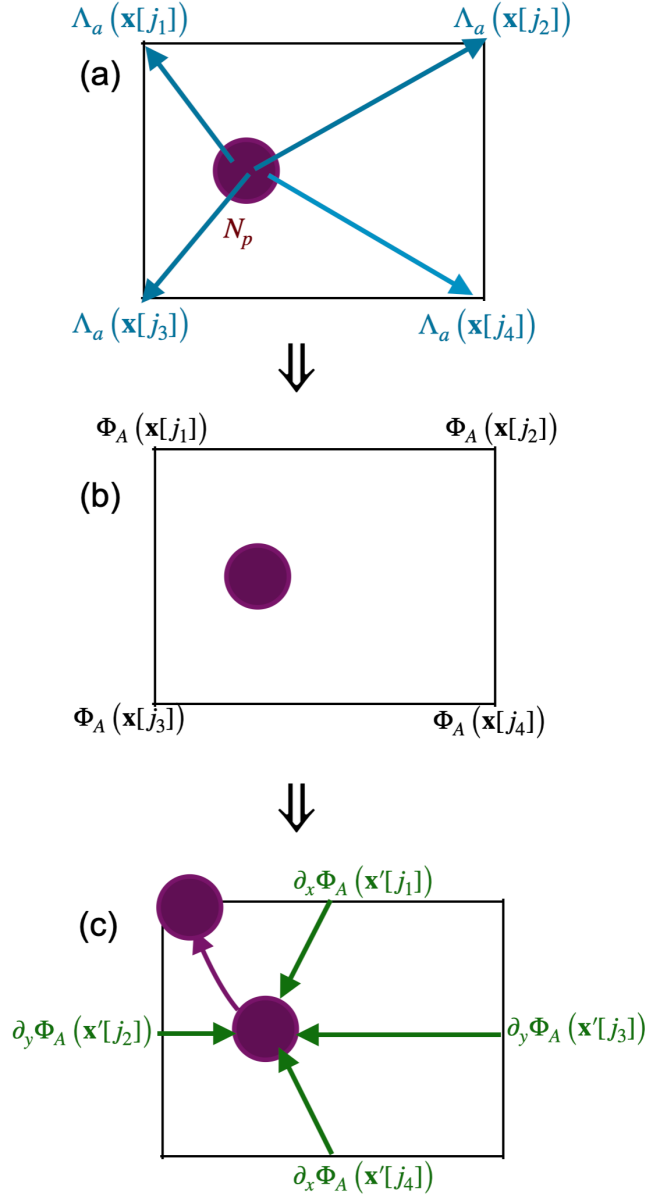}}
\caption{\label{RIC} Schematic of the RIC method. Panel (a) shows the derivation of $\Lambda_a$ on the nodes from the projections of $N_p$ calculated on the rays. Panel (b) shows the derivation of $\Phi_A$ on the nodes from the knowledge of $\Lambda_a$. Panel (c) shows the derivation of $\nabla \Phi_A$ on the cells midpoints and its projection on the rays to derive the rays dynamics.}
\end{figure}
The RIC numerical scheme follows from that introduced in~\rf{deb19}, where the wave quanta are calculated along rays, and the wave amplitudes are estimated at the nodes of parallelipedic cells. The wave amplitudes at a given cell node follow from a simple averaging over all the rays located inside the cell. Consequently, these amplitudes are necessarily bounded from above, the upper bound being fixed by the cell volume, $\Delta V$. This allows to avoid the divergence in amplitude inherent to the use of the geometrical optics approximation when the rays cross each other, e.g., at caustics or when the wave self-focuses. Actually, in the RIC method, we do not use simple averages as in \rf{deb19}. Instead, we first estimate the wave amplitude at the cell nodes, using a shape factor. This allows us to derive the gradient of the field amplitude on the mesh midpoints,  which we project back onto the rays using the same shape factor. This technique is borrowed from PIC codes, whence the acronym RIC. Moreover, unlike in~\rf{deb19}, we use a nonstationary ray tracing, and the fields are derived from a deterministic resolution of Eqs.~(\ref{ray1}), (\ref{ray3}) and (\ref{envd}), and not from a Monte-Carlo method. 

More precisely, from Eq.~(\ref{envd}), we may only derive $J\partial_\omega\chi_aE_A^2$, while the gradient of $\Phi_A=(k\lambda_D)^2E_A/k$  is needed to solve the ray equation~(\ref{ray3}). As a first step to derive $\Phi_A$, we get rid of the Jacobian in order to estimate $\Lambda_a\propto\partial_\omega \chi_aE_A^2$ on the cell node, ${\bm x}_{[j]}$,
\begin{equation}
\label{new1}
\Lambda_a\left({\bm x}_{[j]},t\right)= \int \Lambda_a\left({\bm x},t\right)\delta\left({\bm x}_{[j]}-{\bm x}\right)d{\bm x},
\end{equation}
where $\delta(\bm{x})$ denotes the Dirac distribution. Now, if the variations of $\Lambda_a$ are sufficiently smooth and small over one cell, Eq.~(\ref{new1}) may be replaced by
\begin{equation}
\label{inter1}
\Lambda_a\left({\bm x}_{[j]},t\right)\approx \int \Lambda_a\left({\bm x},t\right)S^{(n)}\left({\bm x}_{[j]}-{\bm x}\right)\frac{d{\bm x}}{\Delta V},
\end{equation}
where $S^{(n)}$ is a shape factor of order $n$~\cite{langdon}. It is such that $\int S^{(n)}d\bm{x}/\Delta V=1$, so that $S^{(n)}$ is dimensionless. From $\Lambda_a\left({\bm x},t\right)=N_p\left[{\bm x}_R(t)\right]/J$, with $J=\vert d{\bm x}_R(t)/{\bm x}_R(0)\vert$, Eq.~(\ref{inter1}) reads
\begin{equation}
\label{inter2}
\Lambda_a\left({\bm x}_{[j]},t\right)\approx \int N_p\left[{\bm x}_R(t)\right]S^{(n)}\left[{\bm x}_{[j]}-{\bm x}_R(t)\right]\frac{d{\bm x}_R(0)}{\Delta V}.
\end{equation}
The estimate of $\Lambda_a$ at the cells nodes, using Eq.~(\ref{inter2}), is illustrated by the panel (a) of Fig.~\ref{RIC} showing a schematic of the RIC method. 

Since we only evaluate $N_p$ over a discrete set of ${\bm x}_R$'s, we replace the integral in Eq.~(\ref{inter2}) by a Riemann sum. In our simulations, we choose the initial ray positions, ${\bm x}_R(0)$, evenly distributed over each cell. The number of initial rays may vary from one cell to the other, which lets us associate an initial volume,  $d{\bm x}_R^i(0)$, to each ray $i$. If ray $i$ starts from a cell where we have placed $N_0^i$ initial positions then, clearly, $d{\bm x}_R^i(0)=\Delta V/N_0^i$. This lets us approximate Eq.~(\ref{inter2}) by
\begin{equation}
\label{inter3}
\Lambda_a\left({\bm x}_{[j]},t\right)\approx\sum_i \frac{N_p\left[{\bm x}_{R_i}(t)\right]}{N_0^i}S^{(n)}\left[{\bm x}_{[j]}-{\bm x}_{R_i}(t)\right].
\end{equation}
For the sake of simplicity, in our simulations, we chose the same number of initial rays in each cell, so that $N_0^i$ is a constant, which we denote by $N_0$. Moreover, all the results presented in this Section have been obtained by using a first-order shape factor. 

From the value of $\Lambda_a$ at ${\bm x}_{[j]}$, we derive that of $\Phi_A$ at the same location by solving
\begin{equation}
\label{phi1}
k^2\Phi_A^2\partial_\omega\chi_a(\Phi_A)=\frac{4\Lambda_a\left({\bm x}_{[j]},t\right)}{(k\lambda_D)^4\varepsilon_0},
\end{equation}
which corresponds to panel (b) of Fig.~\ref{RIC}. Once $\Phi_A$ is known on a regular mesh, its gradient is easily derived on the cells midpoints by making use of finite differences. Then, $\bm{\nabla}\Phi_A$ is projected back onto the rays, using the same shape factor, 
$S^{(n)}$,  as for $\Lambda_a$. This allows to estimate the right-hand side of Eq.~(\ref{ray3}) and to move the rays forward,  as illustrated by the panel (c) of Fig.~\ref{RIC}.

The ray equations~(\ref{ray1})~and~(\ref{ray3}) are solved using the same time step, $\delta t$, as for Eq.~(\ref{envd}) on $N_p$. Therefore, all quantities are always estimated on the same location along a ray. $\delta t$ is chosen so that $\gamma_0\delta t$ be small enough, $\gamma_0\delta t \alt 10^{-2}$. We use a symplectic leap-frog time integrator~\cite{hai10} to solve Eqs.~(\ref{ray1}) and (\ref{ray3}), while $N_p$ is derived from Eq.~(\ref{envd}) the following way,
\begin{equation}
N_p(t+\delta t)=N_p(t)\exp\left\{\left[\gamma_0(t+\delta t)+\gamma_0(t)\right]\delta t\right\},
\end{equation}
where we have denoted $\gamma_0(t+\delta t)\equiv \gamma_0[I_{\rm{las}}(t+\delta t),E_a(t)]$. Moreover, $N_p$ is initialized at the same noise level in all cells, whose value, $N_B$, is discussed in Paragraph~\ref{V.4}. 

The cell sizes should be chosen so that the variations of $N_p$ within each cell be small enough for the estimate Eq.~(\ref{inter2}) to remain accurate. In particular, their transverse size, $l_{\bot}$, should be significantly less than the laser waist, $w_0$. Indeed, the transverse extent of the EPW could be much less than $w_0$ due to its inhomogeneous amplification and to self-focussing. Moreover, $l_{\bot}$ should be at least of the order of the wavelength, $\lambda$. Indeed, if $l_{\bot}\ll \lambda$, Eq.~(\ref{inter2}) overestimates $\Lambda_a$ when most rays are located within a few cells due to self-focussing. Hence, $l_{\bot}$ should be chosen so that $l_{\bot}\approx \lambda$. If the ray direction changes significantly due to the $\mk$-rotation, it is not possible to clearly identify the longitudinal and transverse directions over the whole simulation domain. Then, the cells should be cubes with volume $\Delta V\sim \lambda^d$ ($d$ being the dimension for the simulation). 

Since the rays are moving from left to right, they eventually leave the leftmost cells of the simulation box. Then, in these cells, the gradient of the wave amplitude is ill defined. In order to overcome this difficulty, we create a zone, on the left part of the simulation box, where we replace Eq.~(\ref{envd}) by,
\begin{equation}
\label{left}
d_tN_p=2\gamma N_p,
\end{equation}
and where $\gamma$ linearly rises from 0 to $\gamma_0$, defined by Eq.~(\ref{gamma0}). Moreover, the leftmost part of this zone is fed with rays which move at the linear group velocity and which carry a number of plasmons set to the noise level. Hence, the leftmost cells of our simulation box are never void of rays.

The transverse boundaries of the simulation box also have to be treated with care. Indeed, on a node located away from these boundaries, are projected the plasmon numbers carried by the rays which are below and above it. However, a node located at the upper boundary can only receive the contributions from the rays which are below it. Indeed, there is no ray above the upper boundary of the simulation box. Consequently, the wave amplitude at such nodes could be underestimated, which would entail spurious gradients and lead to a wrong estimate of the rays trajectories. In order to alleviate this difficulty, each node carries a number of plasmons set to the noise level, $N_B$, \textit{before} the projections from the rays to the nodes. Then, instead of projecting the number of plasmons, $N_p$, carried by each ray, we only project $N_p-N_B$. Namely, we replace $N_p$ with $N_p-N_B$ in Eq.~(\ref{inter3}). This would lead to a correct estimate of the EPW amplitude on the mesh, provided that the number of plasmons carried by the rays near the transverse boundaries remain close to $N_B$. Hence, the laser intensity at these boundaries must be so weak that it cannot significantly amplify the plasma wave. 

Because of the EPW self-focussing entailed by wavefront bowing, if the rays move according to geometrical optics, they converge towards the beam axis (chosen as the $x$-axis) and can cross it. However, physically, when a ray gets very close to the $x$-axis, it is reflected back due to diffraction. It cannot cross the axis. Once it starts to be reflected and moves away from the axis, the nonlinear frequency gradient bends its trajectory again and lets it converge back towards the axis. Hence, on the average, this ray moves along the $x$-axis, so that the averaged value of $\bm{k}_\bot$ is 0. In order to qualitatively reproduce this feature, we multiply $k_{y,z}$ by $\tanh(\vert y,z\vert/2\lambda)/\tanh(1)$  for all rays such that $\vert y,z\vert<2\lambda$. However, for the simulation parameters detailed in Paragraph~\ref{V.4}, we usually do not have to do so. Indeed, the EPW usually breaks before the rays could have a chance to cross the $x$-axis. 

Our model stems from the envelope equations derived in Ref.~\onlinecite{benisti18}, which are only valid for nearly monochromatic waves. Consequently, additional modeling is required to correctly describe the EPW once it has broken. The PIC simulation results reported in Ref.~\onlinecite{rousseaux} show that the field amplitude dramatically drops after wave breaking. This is most probably due to electron acceleration by chaotic transport, at the expense of the wave energy. In order to account for it, when the EPW amplitude on a ray is too large, we reduce the number of plasmons carried by this ray. Namely, when $\Phi_A\geq \Phi_{\rm{wb}}$, where $\Phi_{\rm{wb}}$ is close to the upper bound for wave breaking derived in Section~\ref{III}, we only project on the mesh a fraction of the number of plasmons carried by the ray. More precisely, we project a number of plasmons that linearly decreases from $N_p$ to 0 when $\Phi_A$ varies from $\Phi_{\rm{wb}}$ to $1.1\times\Phi_{\rm{wb}}$. Moreover, we remove from the simulation box all the rays such that $\Phi_A>1.1\times\Phi_{\rm{wb}}$. Then, clearly, $1.1\times\Phi_{\rm{wb}}$ must be less than the maximum value for $\Phi_A$ deduced from the results of Section~\ref{III}, which we denote by $\Phi_A^{\max}$. However, one cannot just choose $\Phi_{\rm{wb}}=\Phi_A^{\max}/1.1$ because, due to self-focusing, the local wave amplitude on a node may be larger than the maximum amplitude on the rays. For the RIC simulation results of Paragraph~\ref{comp_pic}, which correspond to $\kld=0.14$, $\Phi_A^{\max}\approx 0.68$, and we choose $\Phi_{\rm{wb}}=0.5$.  Then, in our RIC simulation, the maximum value reached by $\Phi_A$ is close to 0.58, which is in very good agreement with the PIC simulation results of Ref.~\onlinecite{rousseaux}  reproduced in Figs.~\ref{phi_surf} (d)-(f), as further discussed in Paragraph~\ref{setup}.

\subsection{Simulation results}
\label{V.4}
\subsubsection{Simulation setup}
\label{setup}
In our simulations, we assume that the laser propagates along the $x$ direction, in a two-dimensional (2-D) plane geometry, 
$(x,y)$. The intensity distribution is a  Gaussian in space and time, $t$,
\begin{equation}
I_{\rm{las}}(x,y,t) =  I_0 \frac{w_0}{w(x)} \exp \left[ \frac{- 2 y^2}{w(x)^2} \right] 
 \times \exp \left[   \frac{ -( t - t_{\rm del})^2}{ \tau^2/4\ln2}  \right]  \, , \label{e:Ilas}
\end{equation}
where $\tau$ is the FWHM pulse duration, and $t_{\rm del}=s/c$, where $s$ is the curvilinear coordinate along the laser ray. For the simulation results of Paragraph~\ref{comp_pic}, we have chosen $s=0$ when $x\omega_{\rm{las}}/c=500$. Moreover, in Eq.~(\ref{e:Ilas}), $w(x) = w_0 \sqrt{1 + [(x-x_f)/l_{\rm R}]^2}$, where $w_0$ is the laser waist, $x_f$ is the abscissa at best focus, and $l_{\rm R} = \pi w_0^2 / \lambda_{\rm{laser}}$ is the Rayleigh length ($\lambda_{\rm{laser}}$ being the laser wavelength). 

\begin{figure}[!h]
\centerline{\includegraphics[width=12cm]{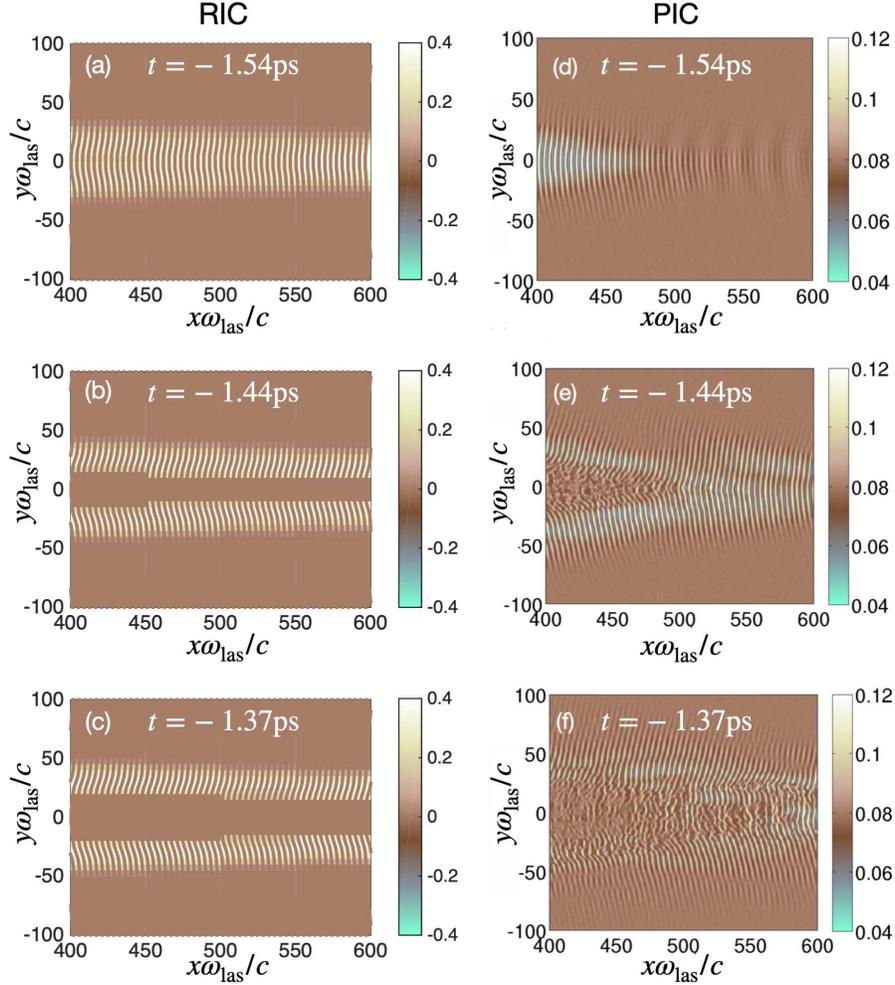}}
\caption{\label{phi_surf} In panels (a)-(c), the colormap indicates the values of $-\delta n_e^{\min}/n_e$ as derived from our RIC simulation, and deduced from $\Phi_A$ by using Eq.~(\ref{dn}) with ten harmonics in the sum. On top of these maps are drawn the curves perpendicular to the local wavenumber, in the same color as that corresponding to $\delta n_e^{\min}=0$. Panels (d)-(f) plot the electron density (normalized to the critical one) as derived from the PIC simulation, and reprinted from C. Rousseaux, S. D. Baton, D. B\'enisti, L. Gremillet, J. C. Adam, A. H\'eron, D. J. Strozzi and F. Amiranoff, (2009), 
``Experimental Evidence of Predominantly Transverse Electron Plasma Waves Driven by Stimulated Raman Scattering of Picosecond Laser Pulses,'' Phys. Rev. Lett. \textbf{102}, 185003:1-4. Panels (a) and (d) are at $t=-1.54\text{ ps}$, panels (b) and (e) at $t=-1.44\text{ ps}$, panels (c) and (f) at $t=-1.37\text{ ps}$.}
\end{figure}

We choose our parameters so as to simulate the experiment  on the LULI 100-TW laser system published in~\rf{rousseaux}. This allows direct comparisons with the 2-D PIC simulations reported on the same publication. 
Hence, we choose $\lambda_{\rm{laser}} = 1060$ nm, $\tau=1.6$ ps, $I_0 = 2.5\times10^{17}$ W/cm$^2$, and $w_0 = 16\lambda_{\rm las}/ \pi$ for an $f/8$ beam aperture. Moreover, the laser is assumed to be focussed at $x_f=1000c/\omega_{\rm{las}}$. As for the simulated plasma, it is homogeneous, with density $n_e/n_c=0.08$ ($n_c=\varepsilon_0m\omega_{\rm{las}}^2/e^2$ being the critical density), and its temperature is 300 eV. Then, the EPW resulting from SRS is such that $\kld\approx 0.14$. 

The simulation box ranges from $y=-100c/\omega_{\rm{las}}$ to $y=100c/\omega_{\rm{las}}$, and from $x=-500c/\omega_{\rm{las}}$ to $x=1000c/\omega_{\rm{las}}$ (it is 253 $\mu$m long and  33.7 $\mu$m wide). When $x\leq 0$, the EPW amplification is derived from Eq.~(\ref{left}) with $\gamma$ varying linearly from 0 to $\gamma_0$, when $x$ varies from $-500c/\omega_{\rm{las}}$ to $0$.  When $x\geq 0$, we solve Eq.~(\ref{envd}) to derive $N_p$. In the plasma domain $400\leq x\omega_{\rm{las}}/c\leq600$ which we investigated more particularly, when $t\leq-1\text{ ps}$, and in the domain $20\leq \vert y\vert\omega_{\rm{las}}/c\leq 50$ where bowing is most effective, the averaged value of $\gamma_0/kv_{th}$ is close to 0.25. This is above the condition for adiabaticity, $\gamma_0\alt0.1$. Note, though, that we do not account for pump depletion, that should make the adiabatic approximation used to derive $\Omega_R$ more accurate. As a matter of fact, and as discussed in Paragraph~\ref{comp_pic}, our results compare very well with those from the PIC simulations of Ref.~\onlinecite{rousseaux}.

The EPW rays are assumed to be initially aligned with the laser rays. Consequently, the initial values of $k_x$ and $k_y$ are derived from the gradient of the complex phase of the Gaussian beam~\cite{siegman}. Moreover, the initial amplitude on each ray corresponds to the noise level, $\Phi_A=5\times10^{-8}$. It has been chosen so that, at $t=-1.54\text{ ps}$ and in the region $400\leq x\omega_{\rm{las}}/c\leq600$, the maximum value of $\Phi_A$ on the rays be close to the limit we choose for wave breaking, $\Phi_{\rm{wb}}=0.5$. Because of self-focusing, the maximum value of $\Phi_A$ on the cell nodes exceeds that carried by the rays. When $t=-1.54\text{ ps}$, this maximum value is close to 0.58. Note that, from Eq.~(\ref{newnew}), $\Phi_A=\sqrt{\sum_jj^2\Phi_j^2}$, while with our normalization and from Poisson equation, the density fluctuation induced by the EPW, $\delta n_e$,  is such that $\delta n_e/n_e=-\sum_jj^2\Phi_j\cos(j\varphi)$. When $\kld=0.14$, the minimum value for $\delta n_e$~\cite{noteF} is reached when $\varphi=0$ so that, 
\begin{equation}
\label{dn}
-\frac{\delta n_e^{\min}}{n_e}=\sum_jj^2\Phi_j.
\end{equation}
Because $\delta n_e^{\min}/n_e$ converges more slowly than the potential, we use then harmonics (instead of three)  to derive it from Eq~(\ref{dn}). Then, for the largest amplitude reached by $\Phi_A$ in our RIC simulation when $t=-1.54\text{ ps}$, $\Phi_A\approx0.58$, we estimate $-\delta n_e^{\min}/n_e\approx 0.38$. This is in very good agreement with the PIC simulation results of Ref.~\onlinecite{rousseaux}. Indeed, as may be inferred from Fig.~\ref{phi_surf} (d)~\cite{noteF}, just before the EPW starts to break, the minimum value reached by the electron density is close $0.05n_c$, so that $-\delta n_e^{\min}/n_e\sim 35-40\%$ (since $n_e/n_c=0.08$). Hence, we choose our noise level so as to match the PIC simulation results at $t=-1.54$ ps as regards the maximum wave amplitude. 

We use the same time step, $\delta t=1$~fs, to numerically solve Eqs.~(\ref{ray1}), (\ref{ray3}) and (\ref{envd}) from $t=-2.2\text{ ps}$ (like in the PIC simulation of Ref.~\onlinecite{rousseaux}). Our mesh is made of rectangular cells, with longitudinal size $l_x=50c/\omega_{\rm{las}}$, and transverse size $l_y=5c/\omega_{\rm{las}}$. Hence, there are only 30 cells along the $x$-direction and 40 ones along the $y$-direction. This very low resolution is enough for the RIC method to yield accurate results (no significant change could be found in our results when $l_x$ and $l_y$ were reduced by a factor of 5). This makes the method very effective. When using 64 rays per cell, the results of Paragraph~\ref{comp_pic} are obtained within a CPU time of 2 minutes. This is about $10^6$ times faster than a PIC simulation.
 
\subsubsection{Results from the RIC simulation}
\label{comp_pic}

In this Paragraph, we present our RIC simulation results regarding wavefront bowing, which we systematically compare against those from the PIC simulation of \rf{rousseaux}. Consequently, whenever we refer to the PIC simulation, we actually mean ``the PIC simulation of \rf{rousseaux}'', without systematically specifying it. \\

Figs.~\ref{phi_surf} (a)-(c) plot the maps of $-\delta n_e^{\min}/n_e$, estimated on the cells nodes from our RIC simulation, and deduced from $\Phi_A$ using Eq.~(\ref{dn}). On top of these maps, we plot the curves perpendicular to the local wavenumber. These curves mimic the wavefronts. They are plotted with the same color as that corresponding to $\delta n_e^{\min}=0$, so that they would not appear where the wave amplitude is very small. Although $-\delta n_e^{\min}\geq0$, we let the colormap showing the values of $-\delta n_e^{\min}/n_e$ go from -0.4 to 0.4, so that $\delta n_e^{\min}=0$ would correspond to the same brown color as in Figs.~\ref{phi_surf} (d)-(f) reproducing the PIC simulation results of Ref.~\onlinecite{rousseaux}. These figures plot the actual density, which oscillates at the local wavelength, allowing a direct visualization of the wavefronts. 

Figs.~\ref{phi_surf} (a) and (d) compare the RIC and PIC simulation results at $t=-1.54$ ps. In both these figures, wavefront bowing is very similar, and the EPW is amplified over the same transverse region, but not over the same longitudinal one.  The EPW is strongly amplified up $x\approx 750c/\omega_{\rm{las}}$ in the RIC simulation (not shown here), and only up to $x\approx 500c/\omega_{\rm{las}}$ in the PIC simulation. Hence, as regards the region where the EPW is strongly amplified, the agreement is not perfect because we do not solve the actual three-wave problem for SRS, and neglect pump depletion. However, wavefront bowing is very well reproduced in our RIC simulation, thus meeting  our prime objective. 

The brown region in the center of Figs.~\ref{phi_surf} (b) and (c), is where we have withdrawn the rays which carry such a large amplitude that we estimate that the EPW is totally broken. As discussed in Paragraph~\ref{V.3}, this happens where $\Phi_A>1.1\Phi_{\rm{wb}}$, where we have chosen $\Phi_{\rm{wb}}=0.5$. At $t=-1.44$ ps, and at $x=400c/\omega_{\rm{mas}}$, we find in our RIC simulation that the EPW is broken over a region that extends up to $\vert y\wl/c\vert\approx20$, as may be seen in Fig.~\ref{phi_surf} (b). This is in good agreement with the PIC simulation result plotted in Fig.~\ref{phi_surf} (d). However, in the PIC simulation the EPW is broken only up to $x\approx 500c/\omega_{\rm{las}}$, while in the RIC simulation it is broken up to $x\approx 700c/\omega_{\rm{las}}$ (not totally shown here), although over only a narrow region $\vert y\wl/c\vert<10$ when $x\wl/c>450$. Hence, there is a fair agreement between the RIC and PIC results as regards wave breaking, although the agreement is not perfect because we neglect pump depletion in the RIC simulation.

At $t=-1.37$ ps, the EPW is broken over about the same region in the RIC and PIC simulations, at least within the domain $400\leq x\omega_{\rm{mas}}/c\leq600$, as may be seen in Figs.~\ref{phi_surf} (c) and (f). This lets us conclude that, not only can the RIC method be used after the EPW has broken, but it also gives a fair account of the space region where the EPW is broken. This is quite remarkable considered the simplicity of the model, compared to the complexity of wave breaking. 

One may also appreciate in Figs.~\ref{phi_surf} (a)-(c) the very low definition that was enough to use in our RIC simulation to get accurate results. This is one of the main reasons for the effectiveness of the RIC method. \\

\begin{figure}[!h]
\centerline{\includegraphics[width=12cm]{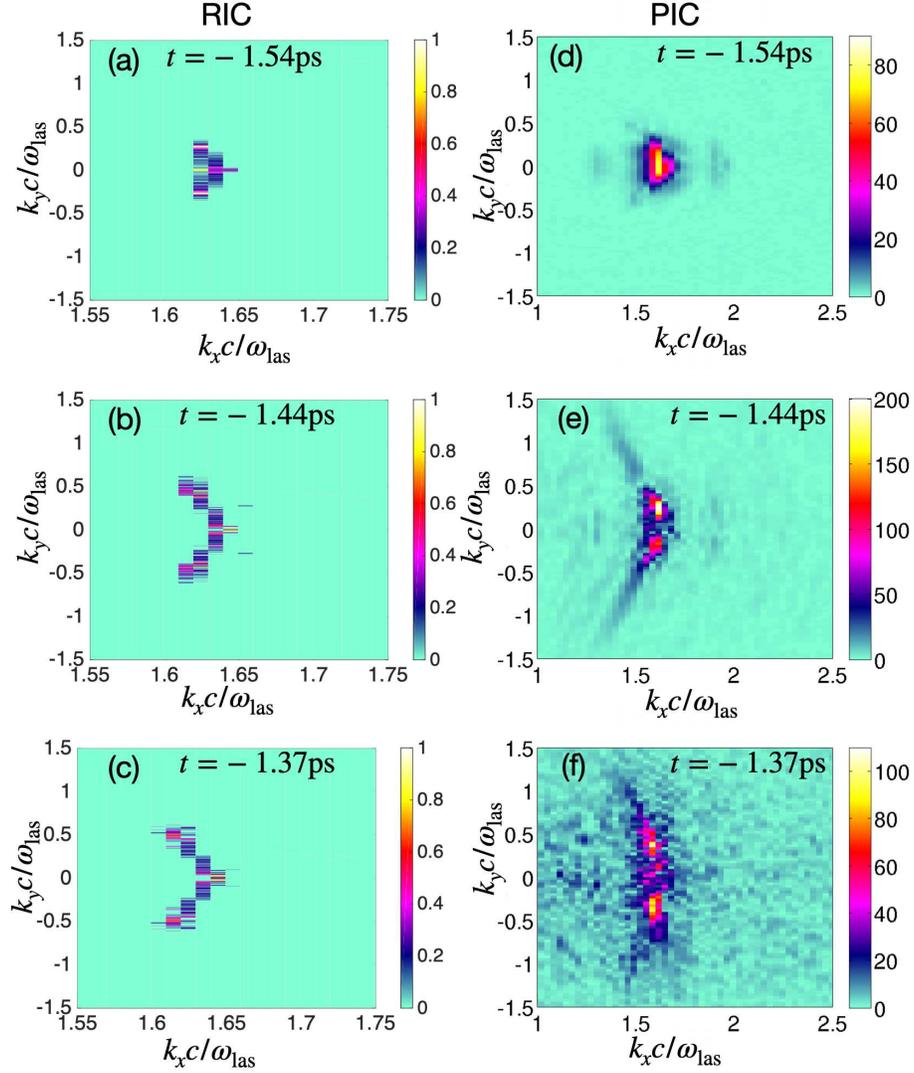}}
\caption{\label{histo} Panels (a)-(c) plot the values of $\langle \tilde{n}_e^2(k_x,k_y)\rangle$ defined by Eq.~(\ref{nk}) and as derived from our RIC simulation, normalized to their maximum value. Panels (d)-(f) plot the modulus squared of the Fourier transforms of the electron density, in arbitrary units, as derived from the PIC simulation and reprinted form C. Rousseaux, S. D. Baton, D. B\'enisti, L. Gremillet, J. C. Adam, A. H\'eron, D. J. Strozzi and F. Amiranoff, (2009), 
``Experimental Evidence of Predominantly Transverse Electron Plasma Waves Driven by Stimulated Raman Scattering of Picosecond Laser Pulses,'' Phys. Rev. Lett. \textbf{102}, 185003:1-4. Panels (a) and (d) are at $t=-1.54\text{ ps}$, panels (b) and (e) at $t=-1.44\text{ ps}$, panels (c) and (f) at $t=-1.37\text{ ps}$.}
\end{figure}

In order to make our comparisons with the PIC simulation more quantitative, we compute
\begin{equation}
\label{nk}
\langle \tilde{n}_e^2(k_x,k_y)\rangle=\sum_i k_i^4\Phi_{A_i}^2,
\end{equation}
where the sum is over all the rays located in the region $400<x\omega_{\rm{laser}}/c<600$, and whose wavenumbers, $\mk_i$, are such that $\vert k_{x,y}-k_{i_x,i_y}\vert<10^{-2}\wl/c$. Figs.~\ref{histo} (a)-(c) show the maps of $\langle \tilde{n}_e^2(k_x,k_y)\rangle$, normalized to its maximum value, at times $t=-1.54\text{ ps}$, $t=-1.44\text{ ps}$ and $t=-1.37\text{ ps}$. Clearly, $\langle \tilde{n}_e^2\rangle$ mimics the square of the Fourier transform of the EPW  density. Hence, the results of Figs.~\ref{histo} (a)-(c) are compared with those of the Fourier transform of the density, derived from the PIC simulation, and reproduced in Figs.~\ref{histo} (d)-(f). At $t=-1.54\text{ ps}$,  the $k_y$-span in $\langle \tilde{n}_e^2\rangle$ found from our RIC simulation is the same as that of the Fourier transform of the EPW density, as derived from the PIC simulation. At this time, the EPW is not broken, so that the extent in $k_y$ is only due to wavefront bowing. This lets us conclude that our RIC simulation does estimate very accurately the transverse modes which only result from bowing. However, the $k_x$-span in the PIC Fourier spectrum is larger than in the $\langle \tilde{n}_e^2\rangle$ map. This is most probably due to the fact that, unlike the   $\langle \tilde{n}_e^2\rangle$ map, the Fourier spectrum accounts for the $x$-variation of the wave amplitude. This entails a width in $k_x$ which is not related to the longitudinal gradient of the EPW frequency, $\Omega_R$.

At $t=-1.44\text{ ps}$, the $k_y$-span in $\langle \tilde{n}_e^2\rangle$ is very similar to that of the PIC Fourier spectrum of the EPW density. This shows that the latter is mainly due to bowing, although the EPW has already broken in the domain $400\leq x\wl/c\leq500$. However, there is a well marked maximum in the Fourier spectrum at $k_y\approx0.25$, and a minimum at $k_y\approx0$, absent from the  $\langle \tilde{n}_e^2\rangle$ map. This suggests that, at  $t=-1.44\text{ ps}$, the density spectrum is affected by the growth of sidebands, especially close to $k_y=0$. 

At $t=-1.37\text{ ps}$, the Fourier spectrum of the density is most significant when $\vert yc/\wl\vert\leq0.6$, which corresponds to the span in $k_y$ for the $\langle \tilde{n}_e^2\rangle$ map in Fig.~\ref{histo} (c). Therefore, even at $t=-1.37\text{ ps}$, when the EPW is broken over a significant part of the space domain, the main features of the density spectrum result from bowing. This PIC spectrum also contains some low signal at large transverse wavenumbers, up to $\vert k_y\wl/c\vert\sim 1$. These are not recovered in our RIC simulation. Therefore, we can unambiguously conclude that they result from transverse instabilities. \\

\begin{figure}[!h]
\centerline{\includegraphics[width=8.6cm]{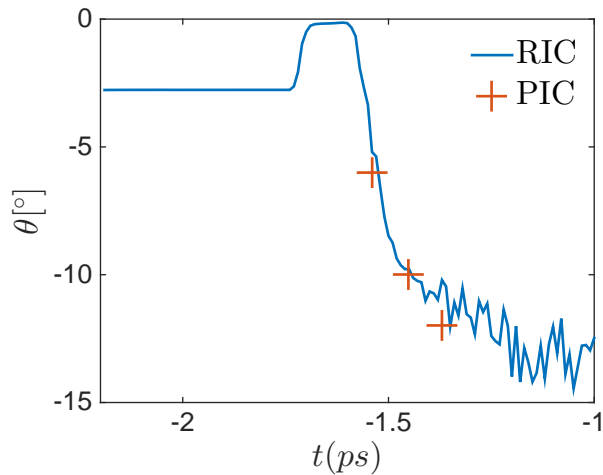}}
\caption{\label{RIC_vs_PIC} Average angle of propagation of the EPW with respect to the $x$-axis, in the upper plane $y>0$, as derived from the PIC simulation (pluses) and from the RIC simulation (solid line).}
\end{figure}

Due to wavefront bowing, the EPW propagates at a nonzero averaged angle, $\theta$, with respect to the averaged direction of propagation of the laser beam (which is the $x$-direction). We estimate $\theta$ from the PIC simulation as in \rf{rousseaux}. At $t=-1.44\text{ ps}$ and $t=-1.37\text{ ps}$, $\theta_{\rm{PIC}}=\tan^{-1}(k_{y_0}/k)$, where $k_{y_0}$ is at the local maximum of the Fourier spectrum. Here, we estimate $k_{y_0}c/\wl\sim \pm(0.25-0.3)$, $\theta_{\rm{PIC}}\approx \pm10$\textdegree{} when  $t=-1.44\text{ ps}$ and $k_{y_0}c/\wl\sim \pm(0.3-0.35)$, $\theta_{\rm{PIC}}\approx \pm12$\textdegree{} when  $t=-1.37\text{ ps}$. When $t=-1.54\text{ ps}$, we use for $k_{y_0}$ the half width at half maximum of the Fourier spectrum, which lets us estimate $k_{y_0}c/\wl\sim \pm(0.15-0.2)$, $\theta_{\rm{PIC}}\approx \pm6$\textdegree{}. These values for $\theta_{\rm{PIC}}$ are reported in Fig.~\ref{RIC_vs_PIC}.

Using our RIC simulation results, we can estimate $\theta$ the following way,
\begin{equation}
\label{75}
\theta_{\rm{RIC}}=\frac{\sum_i\theta_ik_i^4\Phi_{A_i}^2}{\sum_ik_i^4\Phi_{A_i}^2},
\end{equation}
where, for each ray, $\theta_i=\tan^{-1}(k_{y_i}/k_{x_i})$. Moreover, in Eq.~(\ref{75}), the sum is limited to those rays located in the upper plane $y>0$, so that $\theta_{\rm{RIC}}<0$. The values for $\theta_{\rm{RIC}}$ derived from Eq.~(\ref{75}) are plotted in Fig.~\ref{RIC_vs_PIC}. When $t\alt-1.7\text{ ps}$, they are nonzero because of the finite opening angle of the laser beam. The increase in $\theta_{\rm{RIC}}$, and the relatively low values it assumes when $-1.7 \alt t \alt -1.6$, is the consequence of gain narrowing. Indeed, at these times, the EPW is mostly amplified on the rays located close to the $x$-axis, which mainly propagate along the $x$-direction. When $-1.6\alt t\alt -1.2$, $\theta_{\rm{RIC}}$ decreases because of the wavefront bowing. As may be seen in Fig.~\ref{RIC_vs_PIC}, a very good agreement is found between $\theta_{\rm{PIC}}$ and $\theta_{\rm{RIC}}$ at $t=-1.54\text{ ps}$, $t=-1.44\text{ ps}$ and $t=-1.37\text{ ps}$. This shows, once again, the relevance of the RIC method to derive the EPW wavefront bowing. 

After  $t=-1\text{ ps}$, we find that $\theta_{\rm{RIC}}$ starts to increase. This is because the EPW has broken in so large a region that only remain in our simulation the rays located far away from the $x$-axis. There, the laser intensity is so small that the plasma wave is only poorly amplified, and bowing does not really occur. Moreover, when the EPW has broken nearly everywhere, the RIC method becomes doubtful, and the corresponding results are not shown here. \\

The maximum value found for $\vert\theta_{RIC}\vert$ is close to 14\textdegree{}. It is rather small, which vindicates the neglect of the $\bm{k}$-rotation when computing the nonlinear EPW frequency, $\Omega_R$~\cite{benisti20I}. However, this does not mean that the effect of the EPW wavefront bowing is negligible. Indeed, from $\bm{k}_s=\bm{k}_{las}-\bm{k}$, where $\bm{k}_{las}$ and $\bm{k}_s$ are, respectively, the laser and scattered wavenumbers, one finds that the scattered wave would propagate at an average angle close to 35\textdegree{} with respect to the $x$-axis when $\theta\approx14$\textdegree{}. This is a significant angle showing that, due to the nonlinear wavefront bowing, there is an effective side-scattering. This has to be accounted for when modeling laser-plasma experiments. Indeed, this allows to correctly derive where the backscattered light, that may be collected in an experiment, actually comes from. This also allows to correctly account for the effect of SRS on the plasma hydrodynamics.

\section{Conclusion}
\label{VI}
In this paper, we provided a description, as complete as possible, of nonlinear adiabatic electron plasma waves. 

Using the results derived by Dawson~\cite{dawson}, together with the general theory of the companion paper~\cite{benisti20I}, we could find an explicit expression for the electrostatic potential, valid whatever $\kld$ and up to amplitudes close to the wave breaking limit. This expression was for a growing wave, but should remain valid if the wave has not kept on growing, provided that $V_\phi$ has varied more slowly than the width, in velocity, of the separatrix. 

We proved rigorously that an adiabatic EPW could not keep growing beyond an amplitude, $\Phi_1^{\max}$, which we derived.  In practice, the EPW is  expected to break before reaching $\Phi_1^{\max}$, due to the unstable growth of secondary modes. Hence, $\Phi_1^{\max}$ is only an upper bound for the wave breaking limit which, nevertheless, provides a good estimate of the actual one for the physics situation considered in Section~\ref{V}. Moreover, as discussed in Section~\ref{V}, our estimate for $\Phi_1^{\max}$ is useful, and relevant, to address the EPW wavefront bowing. The values we plotted in Fig.~\ref{f5} for the maximum EPW amplitude are only for a growing wave. However, by using the general theory of the companion paper, we can derive $\Phi_1^{\max}$ whatever the space and time variations of the scalar and vector potentials, and of the wavenumber and wave frequency. 

In order to derive the space and time variations of the wave amplitude, one may resort to envelope equations. From Section~\ref{II}, we know that the scalar potential is either sinusoidal, or as derived by Dawson except, maybe, for amplitudes close to the wave breaking limit. Using the latter result, we provided an explicit expression for the nonlinear EPW envelope equation, valid to describe the wave growth up to its breaking, whatever $\kld$. Moreover, we also showed how the general equation could be simplified when the variations of the amplitude of the scalar potential were much faster than those of the vector potential and phase velocity. 

In a multidimensional geometry, the envelope equation needs to be solved along rays, whose trajectories have to be calculated self-consistently while the wave amplitude is changing. We did perform such a nonlinear calculation by using for the EPW an envelope equation that mimicked, in a simplified way, the SRS drive. To the best of our knowledge, this had never been done before, and this let us introduce a new numerical scheme, dubbed ray-in-cell (RIC). From RIC simulations, we could compute transverse modes which only resulted from wavefront bowing. We showed that they compared very well with those derived from the PIC simulations of \rf{rousseaux} before the EPW had broken. This allowed us to unambiguously find which transverse wavenumbers, in the PIC Fourier spectra reproduced in Figs.~\ref{histo} (d)-(f), resulted from bowing or from the growth of secondary instabilities. Moreover, using our RIC simulations, we could estimate the averaged angle of propagation, $\theta$, of the EPW with respect to the $x$-axis, and found that it agreed nicely with that inferred from PIC simulations. Although this angle remains modest, $\theta\alt14$\textdegree{}, it entails a significant angle for the backscattered wave, which may be as large as 35\textdegree{}. Therefore, nonlinear wavefront bowing should induce substantial SRS side-scattering, which has to be accounted for in the modeling of laser-plasma interaction. This is needed in order to correctly estimate the impact of SRS on the plasma hydrodynamics. This is also needed to correctly estimate where the scattered light, that may be collected in an experiment, actually comes from. RIC simulations proved to provide accurate estimates for wavefront bowing at a much reduced computational cost than PIC simulations (they are about $10^6$ faster). They will be used in a future publication to perform large-scale simulations of laser-plasma interaction, accounting for SRS in the nonlinear kinetic regime. 

\begin{acknowledgments}
The authors thank X.~Davoine, C.~Rousseaux, G.~Sary, Y.~Elskens,  and D.~Tordeux for fruitful discussions.
\end{acknowledgments}

\section*{Data availability}

The data that support the findings of this study are available from the corresponding author upon reasonable request.


\begin{thebibliography}{10}
\bibitem{noteD}
One could also think of deriving the nonlinear EPW phase velocity, for Dawson's potential, by solving $1+\chi_a=0$, where $\chi_a$ is defined by Eq.~(\ref{chia}). However, this would not give the correct linear value for the phase velocity, unlike when solving $1+\chi_D=0$, where $\chi_D=2\langle\cos(\varphi)\rangle/\Phi_1$. Consequently, $\chi_D$, and not $\chi_a$, is used in Eq.~(\ref{vg}) to derive $v_g$ with Dawson's potential.
\bibitem{noteE}
Actually, in Ref.~\onlinecite{coffey}, Coffey derives an upper bound for $\mathcal{E}_{\max}=\max\left[\sum_jj\Phi_j\sin(j\varphi)\right]$. We deduced $\Phi_1^{\max}$ from $\mathcal{E}_{\max}$ by multiplying it with $r\equiv\Phi_1/\max\left[\sum_jj\Phi_j\sin(j\varphi)\right]$, as derived from our adiabatic theory. Note, though, that $r$ remains close to unity for all the values of $\kld$ that we investigated (it never departs from unity by more than 6.5\%).
\bibitem{noteF}
The raw data of the PIC simulations, performed more than 10 years ago, are no longer available, and we can only rely on the colormaps to infer the density calculated in these simulations. We focus here on the minimum value for the density, because it can be easily guessed from Fig.~\ref{phi_surf} (d). Using numerical tools that exactly retrieve a pixel's color, we find that the minimum density in Fig~\ref{phi_surf} (d) is very close to 0.05$n_c$. By contrast, the colormap being clearly saturated for the largest values of $n_e$, we cannot guess the maximum density. We only know that $\delta n_e^{\max}/n_e>0.5$, which is in agreement with our theoretical results. Indeed, when $-\delta n^{\min}/n_e\approx0.35$,  we find $\delta n_e^{\max}/n_e\approx 1$.
\end{thebibliography}
\end{document}